\documentclass{vldb}
\usepackage{amsmath}
\usepackage[sort,noadjust,nocompress]{cite}

\usepackage{amsthm}
\usepackage{graphicx}
\usepackage{microtype}
\usepackage{epstopdf}
\usepackage[bf]{caption}
\usepackage[usenames,dvipsnames]{xcolor}
\usepackage{verbatim}
\usepackage{listings}
\usepackage[noend]{algpseudocode}
\usepackage{balance}
\usepackage[normalem]{ulem}

\usepackage{tikz}
\newcommand*\circled[1]{\tikz[baseline=(char.base)]{
            \node[shape=circle,fill=.,inner sep=0pt] (char) {\color{-.}\textsf\footnotesize #1};}}
\def\thepapertitle{Dash: Scalable Hashing on Persistent Memory}

\vldbTitle{\thepapertitle}
\vldbAuthors{Baotong Lu, Xiangpeng Hao, Tianzheng Wang, Eric Lo}
\vldbDOI{https://doi.org/10.14778/3389133.3389134}
\vldbNumber{8}
\vldbVolume{13}
\vldbYear{2020}

\usepackage{mathptmx}
\usepackage{xspace}

\newcommand{\sql}[1]{#1\xspace}
\renewcommand{\texttt}[1]{{\fontfamily{cmtt}\selectfont{#1}}}
\renewcommand{\tt}[1]{{\fontfamily{cmtt}\selectfont{#1}}}

\newcommand{\dash}{\sql{Dash}}
\newcommand{\CAS}{\texttt{CAS}}

\newcommand{\CLWB}{\texttt{CLWB}}
\newcommand{\CLFLUSH}{\texttt{CLFLUSH}}
\newcommand{\CLFLUSHOPT}{\texttt{CLFLUSHOPT}}

\definecolor{comment-red}{rgb}{1,0,0}

\def\thepaperkeywords{Persistent memory, hash tables, concurrency, space utilization, performance}

\usepackage{float}
\usepackage[
  pageanchor=true,
  plainpages=false,
  pdfpagelabels,
  bookmarks,
  bookmarksnumbered,
  pdfborder=0 0 0,  %removes outlines around hyper links in online display
  colorlinks=true,
  linkcolor=blue,
  citecolor=Blue, % OliveGreen
  pdftitle={\thepapertitle},
  pdfauthor={},
  pdfsubject={},
  pdfkeywords={\thepaperkeywords},
]{hyperref}

\usepackage{breakurl}

\floatstyle{ruled}
\usepackage{algorithm}
\newfloat{algorithm}{t}{lop}

\let\oldcite\cite
\renewcommand{\cite}[1]{\colorbox{linkbg}{\oldcite{#1}}}
\definecolor{linkbg}{gray}{0.97}
\newcommand{\myref}[2]{\colorbox{linkbg}{\hyperref[#2]{%
      \ifthenelse{\equal{#1}{}}{}{#1 }\ref{#2}}}}

\makeatletter
\newcommand{\citeme}{\@ifstar
  \citemeStar%
  \citemeNoStar%
}
\def\algbackskip{\hskip-\ALG@thistlm}
\makeatother

\newcommand{\citemeStar}{\citemeNoStar{citation needed}}
\newcommand{\citemeNoStar}[1]{
  {\color{Magenta}{\bf \fbox{citeme} #1}}
}

\usepackage{booktabs}

\usepackage{nicefrac}

\usepackage{paralist}

\usepackage{subcaption}
\usepackage{soul}
\usepackage{multirow}

\usepackage{enumitem}
\usepackage{tabularx}

\begin{document}

\newcolumntype{L}[1]{>{\raggedright\arraybackslash}p{#1}}
\newcolumntype{C}[1]{>{\centering\arraybackslash}p{#1}}
\newcolumntype{R}[1]{>{\raggedleft\arraybackslash}p{#1}}

\newtheorem{definition}{Definition}[]
\newtheoremstyle{defstyle}
  {0em} % Space above
  {0em} % Space below
  {} % Body font
  {} % Indent amount
  {\bfseries} % Theorem head font
  {.} % Punctuation after theorem head
  {0em} % Space after theorem head
  {} % Theorem head spec (can be left empty, meaning `normal')

\setlength{\pdfpageheight}{\paperheight}
\setlength{\pdfpagewidth}{\paperwidth}

\title{\thepapertitle}

\numberofauthors{1}
\author{
\alignauthor
Baotong Lu$^1$\thanks{Work partially performed while at Simon Fraser University.}
~~~
Xiangpeng Hao$^2$
~~~
Tianzheng Wang$^2$
~~~
Eric Lo$^1$
\bigskip\\\vspace{-3mm}
\begin{tabular}{cc}
$^1$\affaddr{The Chinese University of Hong Kong} & $^2$\affaddr{Simon Fraser University}\\
\fontsize{10}{10}\selectfont\sf~\{btlu, ericlo\}@cse.cuhk.edu.hk & \fontsize{10}{10}\selectfont\sf~\{xha62, tzwang\}@sfu.ca
\end{tabular}
\\
}

\newcommand{\BT}[1]{\textcolor{red}{BT: #1}}
\newcommand{\BTD}[1]{\textcolor{blue}{BT: #1}}

\maketitle

\begin{abstract}
Byte-addressable persistent memory (PM) brings hash tables the potential of low latency, cheap persistence and instant recovery. 
The recent advent of Intel Optane DC Persistent Memory Modules (DCPMM) further accelerates this trend. 
Many new hash table designs have been proposed, but most of them were based on emulation and perform sub-optimally on real PM.
They were also piece-wise and partial solutions that side-step many important properties, in particular good scalability, high load factor and instant recovery.

We present \dash, a holistic approach to building dynamic and scalable hash tables on real PM hardware with all the aforementioned properties.
%\dash achieves (1) scalability, (2) high load factor, and (3) instant recovery.
Based on \dash, we adapted two popular dynamic hashing schemes (extendible hashing and linear hashing). 
%\textcolor{BrickRed}{
On a 24-core machine with Intel Optane DCPMM, we show that compared to state-of-the-art, \dash-enabled hash tables can achieve up to $\sim$3.9$\times$ higher performance with up to over 90\% load factor and an instant recovery time of 57ms regardless of data size.
%}
\end{abstract}

\section{Introduction}
\label{sec:intro}
Dynamic hash tables that can grow and shrink as needed at runtime are a fundamental building block of many data-intensive systems, such as database systems~\cite{MySQL,PostgreSQL,MeetWalkers,Cicada,MMDB-Impl,MMDBMS} and key-value stores~\cite{Bluecache,MemC3,SLM-DB,WiscKey,LevelDB,Redis,FASTER}.
Persistent memory (PM) such as 3D XPoint~\cite{Intel3DXP} and memristor~\cite{Memristor} promises byte-addressability, persistence, high capacity, low cost and high performance, all on the memory bus.
%Data in PM can be directly accessed using \texttt{load} and \texttt{store} instructions with persistence but without involving the (expensive) storage stack.
These features make PM very attractive for building dynamic hash tables that persist and operate directly on PM, with high performance and instant recovery.
The recent release of Intel Optane DC Persistent Memory Module (DCPMM) brings this vision closer to reality.
Since PM exhibits several distinct properties (e.g., asymmetric read/write speeds and higher latency); blindly applying prior disk or DRAM based approaches~\cite{ExtHashing,LarsonLinearHashing,LitwinLinearHashing} would not reap its full benefits, necessitating a departure from conventional designs.

%Dynamic hash tables that can grow and shrink at runtime as needed are a fundamental building block of many data-intensive systems, such as database systems~\cite{MySQL,PostgreSQL,MeetWalkers,Cicada,MMDB-Impl,MMDBMS} and key-value stores~\cite{Bluecache,MemC3,SLM-DB,WiscKey,LevelDB,Redis}.
%PM opens up the potential of building hash tables that persist and operate directly on PM, with near-DRAM performance but lower cost of ownership, and instant recovery.
%The recent release of the first (and only) next-generation %\footnote{We focus on scalable PM represented by Intel Optane DC Persistent Memory, instead of traditional DRAM-based NVDIMMs~\cite{AgigaNVDIMM,VikingNVDIMM}.} 
%PM, Intel Optane DC Persistent Memory Module (DCPMM)~\cite{Intel3DXP}, brings this vision closer to reality.

\subsection{Hashing on PM: Not What You Assumed!}
There have been a new breed of hash tables specifically designed for PM~\cite{CCEH,LevelHashing,Dali,PFHT,PathHashing,NVCHashmap} based on DRAM emulation, before actual PM was available.
Their main focus is to reduce cacheline flushes and PM writes for scalable performance.
But when they are deployed on real Optane DCPMM, we find (1) scalability is still a major issue, and (2) desirable properties are often traded off.

\begin{figure}[t]
\centering
\includegraphics[width=1.015\columnwidth]{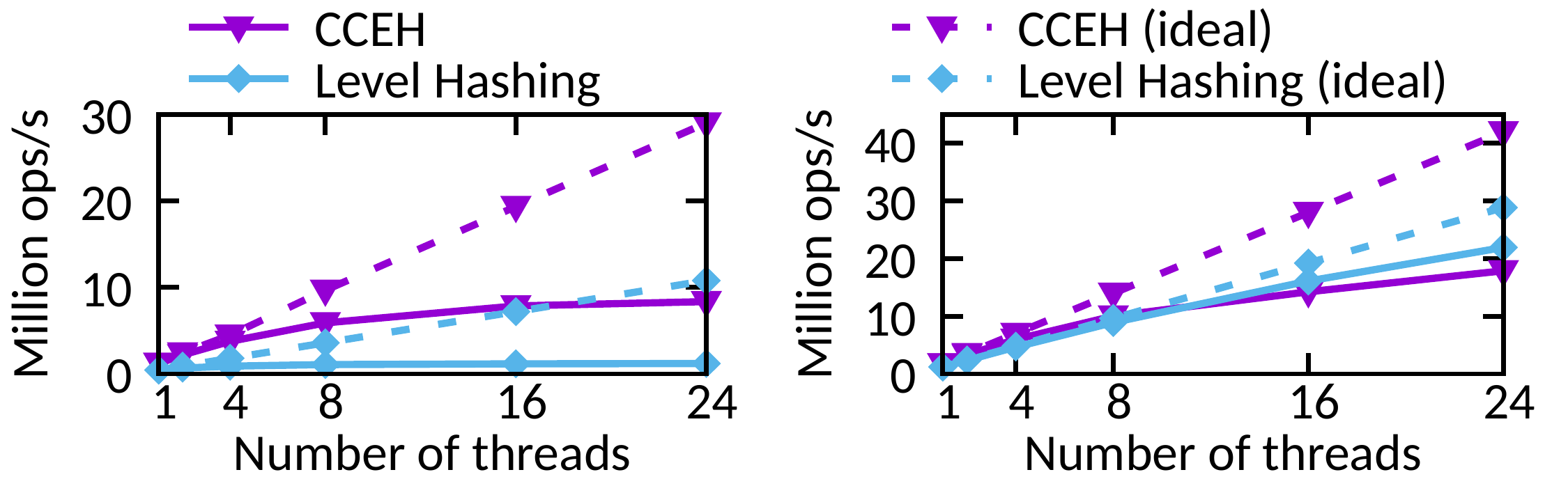}
%\vspace{-3mm}
\caption{Throughput of state-of-the-art PM hashing (CCEH~\cite{CCEH} and
Level Hashing~\cite{LevelHashing}) for insert (left) and
search (right) operations on Optane DCPMM. Neither matches the expected scalability.}
\label{fig:scale-vs-ideal}
\end{figure}

Figure~\ref{fig:scale-vs-ideal} shows the throughput of two state-of-the-art PM hash tables~\cite{LevelHashing,CCEH} under insert (left) and search (right) operations, on a 24-core server with Optane DCPMM running workloads under uniform key distribution (details in Section~\ref{sec:eval}).
As core count increases, neither scheme scales for inserts, nor even read-only search operations.
Corroborating with recent work~\cite{PMPrimitives,PiBench}, we find the main culprit is Optane DCPMM's limited bandwidth which is $\sim$3--14$\times$ lower than DRAM's~\cite{UCSDMeasurement}.
Although the server is fully populated to provide the maximum possible bandwidth, excessive PM accesses can still easily saturate the system and prevent the system from scaling.
We describe two major sources of excessive PM accesses that were not given enough attention before, followed by a discussion of important but missing functionality in prior work.

\textbf{Excessive PM Reads.} Much prior work focused on reducing writes to PM, however, we note that it is also imperative to reduce PM reads; yet many existing solutions reduce PM writes by incurring more PM reads.
Different from the device-level behavior (PM reads being faster than writes), \textit{end-to-end} write latency (i.e., the entire data path including CPU caches and write buffers in the memory controller) is often lower than reads~\cite{PMPrimitives,UCSDGuide}.
The reason is while PM writes can leverage write buffers, PM reads mostly touch the PM media due to hash table's inherent random access patterns.
In particular, existence checks in record probing constitute a large proportion of such PM reads: to find out if a key exists, one or multiple buckets (e.g., with linear probing) have to be searched, incurring many cache misses and PM reads when comparing keys.
%Record probing is ubiquitous in hash tables, required by not only search but also insert and delete operations.
%We identify two major sources of excessive PM accesses in existing hash tables: record probing and concurrency control.

\textbf{Heavyweight Concurrency Control.} Most prior work side-stepped the impact of concurrency control.
Bucket-level locking has been widely used~\cite{CCEH,LevelHashing}, but it incurs additional PM writes to acquire/release read locks, further pushing bandwidth consumption towards the limit. 
Lock-free designs~\cite{NVCHashmap} can avoid PM writes for read-only probing operations, but are notoriously hard to get right, more so in PM for safe persistence~\cite{PMwCAS}.

Neither record probing nor concurrency control typically prevents a well-designed hash table to scale on DRAM.
However, on PM they can easily exhaust PM's limited bandwidth.
These issues call for new designs that can reduce unnecessary PM reads during probing and lightweight concurrency control that further reduces PM writes.

\textbf{Missing Functionality.}
%In addition to scalability bottlenecks, 
We observe in prior designs, necessary functionality is often traded off for performance (though scalability is still an issue on real PM).
(1) Indexes could occupy more than 50\% of memory capacity~\cite{HybridIndex}, so it is critical to improve load factor (records stored vs. hash table capacity).
Yet high load factor is often sacrificed by organizing buckets using larger segments in exchange for smaller directories (fewer cache misses)~\cite{CCEH}. 
As we describe later, this in turn can trigger more pre-mature splits and incur even more PM accesses, impacting performance.
(2) Variable-length keys are widely used in reality, but prior approaches rarely discuss how to efficiently support it. 
(3) Instant recovery is a unique, desirable feature that could be provided by PM, but is often omitted in prior work which requires a linear scan of the metadata whose size scales with data size. 
(4) Prior designs also often side-step the PM programming issues (e.g., PM allocation), which impact the proposed solution's scalability and adoption in reality.

%In summary, prior designs provide only piece-wise, partial solutions, exhibiting limits in both performance and functionality on real PM hardware.
%It is necessary to revisit these issues and devise PM hash tables that scales on real PM hardware without scrificing desirable properties, which is the focus of this paper.

\subsection{Dash}
We present \textit{\dash}, a holistic approach to \underline{d}ynamic \underline{a}nd \underline{s}calable \underline{h}ashing on real PM without trading off desirable properties.
\dash uses a combination of new and existing techniques that are carefully engineered to achieve this goal.
\circled{\textsf{1}} 
We adopt fingerprinting~\cite{FPTree} that was used in PM tree structures to avoid unnecessary PM reads during record probing.
The idea is to generate fingerprints (one-byte hashes) of keys and place them compactly to summarize the possible existence of keys.
This allows a thread to tell if a key possibly exists by scanning the fingerprints which are much smaller than the actual keys.
%The thread then only needs to access the matched fingerprints.
\circled{\textsf{2}} 
Instead of traditional bucket-level locking, \dash uses an optimistic, lightweight flavor of it that relies on verification to detect conflicts, rather than (expensive) shared locks.
This allows \dash to avoid PM writes for search operations.
With fingerprinting and optimistic concurrency, \dash avoids both unnecessary reads and writes, saving PM bandwidth and allowing \dash to scale well. 
\circled{\textsf{3}} \dash retains desirable properties. 
We propose a new load balancing strategy to postpone segment splits with improved space utilization. 
To support instant recovery, we limit the amount of work to be done upon recovery to be small (reading and possibly writing a one-byte counter), and amortize recovery work to runtime.
Compared to prior work that handles PM programming issues in ad hoc ways, \dash uses PM programming models (PMDK~\cite{PMDK}, one of the most popular PM libraries) to systematically handle crash consistency, PM allocation and achieve instant recovery.

Although these techniques are not all new, \dash is the first to integrate them for building hash tables that scale without sacrificing features on real PM.
%\textcolor{BrickRed}{
Techniques in \dash can be applied to various static and dynamic hashing schemes. %, including both static and dynamic hashing. 
Compared to static hashing, dynamic hashing can adjust hash table size on demand without full-table rehashing which may block concurrent queries and significantly limit performance. 
%This is very desirable as full-table rehashing may block concurrent queries, significantly limiting performance. 
In this paper, we focus on dynamic hashing and apply \dash to two classic approaches: extendible hashing~\cite{ExtHashing,CCEH} and linear hashing~\cite{LitwinLinearHashing,LarsonLinearHashing}.
They are both widely used in database and storage systems, such as Oracle ZFS~\cite{ZFS}, IBM GPFS~\cite{GPFS}, Berkeley DB~\cite{BDB} and SQL Server Hekaton~\cite{hekaton}.
%}

Evaluation using a 24-core Intel Xeon Scalable CPU and 1.5TB of Optane DCPMM shows that \dash can deliver high performance, good scalability, high space utilization and instant recovery with a constant recovery time of 57ms. 
Compared to the aforementioned state-of-the-art~\cite{CCEH,LevelHashing}, \dash achieves up to $\sim$3.9$\times$ better performance on realistic workloads, and up to over 90\% load factor with high space utilization and the ability to instantly recover.

\subsection{Contributions and Paper Organization}
We make four contributions. 
%\begin{itemize}[leftmargin=*]\setlength\itemsep{0em}
%\item
First, we identify the mismatch between existing and desirable PM hashing approaches, and highlight the new challenges.
Second, we propose \dash, a holistic approach to building scalable hash tables on real PM. \dash consists of a set of useful and general building blocks applicable to different hash table designs.
Third, we present \dash-enabled extendible hashing and linear hashing, two classic and widely used dynamic hashing schemes.
Finally, we provide a comprehensive empirical evaluation of \dash and existing PM hash tables to pinpoint and validate the important design decisions. 
Our implementation is open-source at: \url{https://github.com/baotonglu/dash}.
%\end{itemize}

In the rest of the paper, we give necessary background in Section~\ref{sec:bg}. %and discuss state-of-the-art and new challenges in Section~\ref{sec:bg}.
Sections~\ref{sec:principles}--\ref{sec:dash-lh} present our design principles and \dash-enabled extendible hashing and linear hashing.
Section~\ref{sec:eval} evaluates \dash.
We discuss related work in Section~\ref{sec:related-work} and conclude in Section~\ref{sec:conclusion}.

\section{Background}
\label{sec:bg}
We first give necessary background on PM (Optane DCPMM) and dynamic hashing, %(extendible~\cite{ExtHashing} and linear~\cite{LarsonLinearHashing,LitwinLinearHashing}) hashing. 
then discuss issues in prior PM hash tables.

\subsection{Intel Optane DC Persistent Memory}
\label{subsec:optane}

\textbf{Hardware.}
We target Optane DCPMMs (in DIMM form factor).
In addition to byte-addressability and persistence, DCPMM offers high capacity (128/256/512GB per DIMM) at a price lower than DRAM's. %; at most 3TB of PM can be installed per socket.
It supports modes: \textit{Memory} and \textit{AppDirect}.
The former presents capacious but slower volatile memory. %; data is erased across reboot cycles without persistence.
DRAM is used as a cache to hide PM's higher latency, with hardware-controlled caching policy.
The AppDirect mode allows software to explicitly access DRAM and PM with persistence in PM, without implicit caching.
Applications need to make judicious use of DRAM and PM.
Similar to other work~\cite{LevelHashing,CCEH,Dali,NVCHashmap}, we leverage the AppDirect mode, as it provides more flexibility and persistence guarantees.

\textbf{System Architecture.}
The current generation of DCPMM requires the system be equipped with DRAM to function properly.
We also assume such setup with PM and DRAM, both of which are behind multiple levels of \textit{volatile} CPU caches.
Data is not guaranteed to be persisted in PM until a cacheline flush instruction (\CLFLUSH{}, \CLFLUSHOPT{} or \CLWB{})~\cite{IntelManual} is executed or other events that implicitly cause cacheline flush occur.
Writes to PM may also be reordered, requiring fences to avoid undesirable reordering.
The application (hash table in our case) must explicitly issue fences and cacheline flushes to ensure correctness.
\CLFLUSH{} and \CLFLUSHOPT{} will evict the cacheline that is being flushed,
while \CLWB{} does not (thus may give better performance). %\footnote{Our tests on Intel Xeon Gold 6252~\cite{XeonGold6252}, however, showed no noticeable performance difference between the \CLWB{} and \CLFLUSHOPT{}.
%We expect \CLWB{} to perform better in future generations.}
After a cacheline of data is flushed, it will reach the asynchronous DRAM refresh (ADR) domain which includes a write buffer and a write pending queue with persistence guarantees upon power failure. 
Once data is in the ADR domain, it is considered as persisted.
Although DCPMM supports 8-byte atomic writes, internally it uses the 256-byte blocks.
But software should not (be hardcoded to) depend on this as it is an internal parameter of the hardware which may change in future generations.

\textbf{Performance Characteristics.}
At the device level, as many previous studies have shown, PM exhibits asymmetric read and write latency, with writes being slower.
It exhibits $\sim$300ns read latency, $\sim$4$\times$ longer than DRAM's.
More recent studies~\cite{PMPrimitives, UCSDGuide}, however revealed that on Optane DCPMM, read latency \textit{as seen by the software} is often higher than write latency. 
This is attributed to the fact that writes (\texttt{store} instructions) commit once the data reaches the ADR domain at the memory controller rather than when reaching DCPMM media.
On the contrary, a read operation often needs to touch the actual media unless the data being accessed is cache-resident (which is rare especially in data structures with inherent randomness, e.g., hash tables).
Tests also showed that the bandwidth of DCPMM depends on many factors of the workload.
In general, compared to DRAM, it exhibits $\sim$3$\times$/$\sim$8$\times$ slower sequential/random read bandwidth. 
The numbers for sequential/random write are $\sim$11$\times$/$\sim$14$\times$. %, respectively.
Notably, the performance of small random stores is severely limited and non-scalable~\cite{UCSDGuide}, which, however, is the inherent access pattern in hash tables. 
These properties exhibit a stark contrast to prior estimates about PM performance~\cite{NVM-DLog,NVRAMEra,vanRenen2018}, and lead to significantly lower performance of many prior designs on DCPMM than originally reported.
%Moreover, recent studies~\cite{UCSDMeasurement,PMPrimitives} show that end-to-end read latency \textit{as seen by software} is often higher than write latency.
%This is because writes to PM are considered persisted once they reach the write buffer in ADR, but reads from PM often have to access the actual device, especially so for hash tables which exhibit randome access patterns~\cite{UCSDMeasurement}.
Thus, it is important to reduce \textit{both} PM reads and writes for higher performance.
More details on raw DCPMM device performance can be found elsewhere~\cite{UCSDMeasurement}; we focus on the end-to-end performance of hash tables on PM.

\subsection{Dynamic Hashing}
\label{sec:bg-dyn-hash}
%\textbf{Static vs. Dynamic Hashing.}
%Dynamic hashing allows buckets to be allocated and deallocated on demand without full-rehashing, while would be required by a static hash table if its size changes.
%Full rehashing will cause excessive amounts of writes to PM and degrade performance, so dynamic hashing is more desirable for PM.

Now we give an overview %of two major dynamic hashing approaches: extendible hashing~\cite{ExtHashing} and linear hashing~\cite{LarsonLinearHashing,LitwinLinearHashing}.
of extendible hashing~\cite{ExtHashing} and linear hashing~\cite{LarsonLinearHashing,LitwinLinearHashing}.
We focus on their memory-friendly versions which PM-adapted hash tables were based upon.
\dash can also be applied to other approaches 
%(e.g., split-ordered lists~\cite{SplitOrderedLists}) 
which we defer to future work.

\begin{figure}[t]
\centering
\includegraphics[width=0.9\columnwidth]{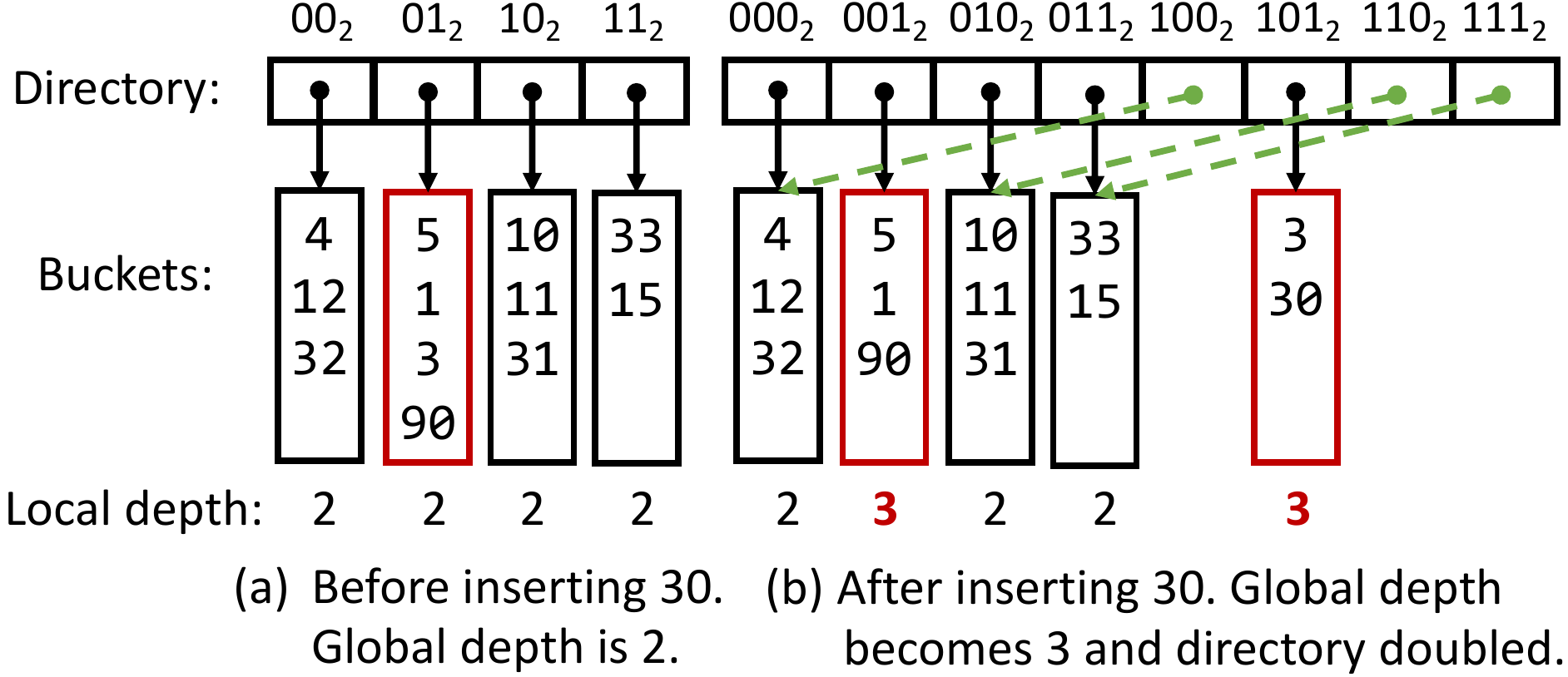}
%\vspace{-3mm}
\caption{\label{fig:exthash} An example of extendible hashing.
%Buckets are indexed by a directory 
%by the most significant \textit{global depth} bits of the hash value.
\textbf{(a)} The hash table is full. \textbf{(b)}
The local depth of unsplit buckets is 
2. Splitting buckets with local depth $<$ global
depth will not double the directory. 
}
\end{figure}

\textbf{Extendible Hashing.}
The crux of extendible hashing is its use of a directory to index buckets so that they can be added and removed dynamically at runtime.
When a bucket is full, it is split into two new buckets with keys redistributed.
The directory may get expanded (doubled) if there is not enough space to store pointers to the new bucket. %; we describe this process later.
Figure~\ref{fig:exthash}(a) shows an example with four buckets, each of which is pointed to by a directory entry; a bucket can store up to four records (key-value pairs).
In the figure, indices of directory entries are shown in binary. % representation. 
The two least significant bits (LSBs) of the hash value are used to select a bucket; we call the number of suffix bits being used here the \textit{global depth}.
%The two most significant bits (MSBs) of the hash value are used to select a bucket for the key; we call the number of prefix bits being used here the \textit{global depth}.
%Note that one can also use the least significant bits of the hash value; we use MSB and describe the impact on PM later.
%Note that one can also use the most significant bits (MSBs) of the hash value; we use MSB and describe the impact on PM later.
The hash table can have at most 2$^{global~depth}$ directory entries (buckets).
A search operation follows the pointer in the corresponding directory entry to probe the bucket.
Each bucket also has a \textit{local depth}. % (described later).
In Figure~\ref{fig:exthash}(a), the local depth of each bucket is 2, same as the global depth.
Suppose we want to insert key 30 that is hashed to bucket 01$_2$, which is full and needs to be split to accommodate the new key.\footnote{Choosing a proper hash function that evenly distributes keys to all buckets is an important but orthogonal problem to our work.}
% tz: this should go to experiments
%Similar to other work, we assume hash values are uniformly distributed.} 
Splitting the bucket will require more directory entries. % so that the new bucket can be accessed.
In extendible hashing, the directory always grows by doubling its current size. 
The result is shown in Figure~\ref{fig:exthash}(b).
Here, bucket 01$_2$ in Figure~\ref{fig:exthash}(a) is split into two new buckets (001$_2$ and 101$_2$), one occupies the original directory entry, and the other occupies the second entry in the newly added half of the directory.
%Here, bucket 01$_2$ in Figure~\ref{fig:exthash}(a) is split into two new buckets (010$_2$ and 011$_2$).
%The newly added entries (green pointers in  Figure~\ref{fig:exthash}(b)) are adjacent to their original entries, pointing to the new bucket or their  corresponding original, unsplit buckets.  
Other new entries still point to their corresponding original buckets.
Search operations will use three bits to determine the directory entry index (global depth now is three).

After a bucket is split, we increment its local depth by one, and update the new bucket's local depth to be the same (3 in our example).
The other unsplit buckets' local depth remains 2.
This allows us to determine whether a directory doubling is needed: if a bucket whose local depth equals the global depth is split (e.g., bucket 001$_2$ or 101$_2$), then the directory needs to be doubled to accommodate the new bucket. 
%This allows us to determine whether a directory doubling is needed: if a bucket whose local depth equals the global depth is split (e.g., bucket 010$_2$ or 011$_2$), then the directory needs to be doubled to accommodate the new bucket.
%\sout{
%Otherwise (local depth $<$ global depth), the directory should already have an entry to accommodate the new bucket, making doubling unnecessary.
%}
%{
%\color{red}
Otherwise (local depth $<$ global depth), the directory should have 2$^{global~depth - local~depth}$ directory entries pointing to that bucket, which can be used to accommodate the new bucket. %, making doubling unecessary. 
%}
%For instance, if bucket 000$_2$ needs to be split, directory entry 100$_2$ that was pointing to bucket 000$_2$ can be updated to point to the new bucket.
For instance, if bucket 000$_2$ needs to be split, directory entry 100$_2$ (pointing to bucket 000$_2$) can be updated to point to the new bucket.

\textbf{Linear Hashing.}
In-memory linear hashing takes a similar approach to organizing buckets using a directory with entries pointing to individual buckets~\cite{LarsonLinearHashing}.
The main difference compared to extendible hashing is that in linear hashing, the bucket to be split is chosen ``linearly.''
That is, it keeps a pointer (page ID or address) to the bucket to be split next and only that bucket would be split in each round, and advances the pointer to the next bucket when the split of the current bucket is finished.
Therefore, the bucket being split is not necessarily the same as the bucket that is full as a result of inserts, and eventually the overflowed bucket will be split and have its keys redistributed.
If a bucket is full and an insert is requested to it, more overflow buckets will be created and chained together with the original, full bucket.
For correct addressing and lookup, linear hashing uses a group of hash functions $h_1...h_n$, where $h_n$ covers twice the range of $h_{n-1}$.
For buckets that are already split, $h_n$ is used so we can address buckets in the new hash table capacity range, and for the other unsplit buckets we use $h_{n-1}$ to find the desired bucket.
After all buckets are split (a \textit{round} of splitting has finished), the hash table's capacity will be doubled; the pointer to the next-to-be-split bucket is reset to the first bucket for the next round of splitting.

Determining when to trigger a split is an important problem in linear hashing.
A typical approach is to monitor and keep the load factor bounded~\cite{LarsonLinearHashing}. %(defined as the number of elements stored in the hash table divided by the capacity of the hash table) bounded~\cite{LarsonLinearHashing}.
The choice of a proper splitting strategy may also vary across workloads, and is orthogonal to the design of \dash.

%\textbf{Other Approaches.}
%\todo{Cuckoo, split-ordered lists?}

\subsection{Dynamic Hashing on PM}
\label{subsec:pm-hash}
%Bucketized cuckoo hashing~\cite{MemC3,libcuckoo} is a widely-used hashing scheme on DRAM because it could guarantee constant lookup time and high memory efficiency. 
%However, it incurs lots of writes due to the excessive cuckoo displacements. 
%Level hashing~\cite{LevelHashing} is a PM-based static hashing scheme that is designed to reduce PM writes and still maintain high memory efficiency. 
%It is a two-level structure that the top-level hash table and bottom-level hash table are both the variant of cuckoo hashing that only allows one cuckoo displacement, to avoid cascading PM writes.  
%To improve the load factor, the bottom-level hash table is not addressable and used to store the records failing to be inserted into the top-level hash table. 
%However, level hashing trades off the read performance because now the search operation needs to access four buckets (top two buckets + bottom two buckets) in the worst case, but cuckoo hashing only needs to probe at most two buckets. 
%To avoid unscalable and expensive full-table rehashing on PM, CCEH~\cite{CCEH} is proposed to support dynamic expansion and shrinkage in a failure-atomic manner. 
%Here we briefly introduce the basic architecture and design, and defer details to Sections~\ref{sec:principles}--\ref{sec:dash-lh}.
Now we discuss how dynamic hashing is adapted to work on PM. 
We focus on extendible hashing and start with CCEH~\cite{CCEH}, a recent representative design; Section~\ref{sec:dash-lh} covers linear hashing on PM.
%We leave other approachs (e.g., split-ordered lists~\cite{SplitOrderedLists}) for future work.

%Existing approaches focused on solving two major problems: (1) excessive PM accesses (reads and writes) and (2) crash consistency (recovery).
%Similar to their DRAM counterparts, extendible hashing on PM uses directories to index buckets for dynamic growth and shrinking.
%However, it 
To reduce PM accesses, CCEH groups buckets into \textit{segments}, similar to in-memory linear hashing~\cite{LarsonLinearHashing}.
Each directory entry then points to a segment which consists of a fixed number of buckets indexed by additional bits in the hash values.
%A certain number of bits in the hash value is used to index buckets and another set of bits is used to index buckets within each segment. 
%The size of the segment is tunable, and in CCEH the default size is 16KB.  
%Each segment typically contains more than 64 buckets, each of which stores metadata and key-value pairs.
%The main benefit of segmentation is cache efficiency and reduced PM accesses: 
By combining multiple buckets into a larger segment, the directory can become significantly smaller as fewer bits are needed to address segments, making it more likely to be cached entirely by the CPU, which helps reducing access to PM.
Note that split now happens at the segment (instead of bucket) level. %, instead of bucket level. %, making it vulnerable to skews.
A segment is split once any bucket in it is full, even if the other buckets in the segment still have free slots, which results in low load factor and more PM accesses.
To reduce such pre-mature splits, linear probing can be used to allow a record to be inserted into a neighbor bucket.
However, this improves load factor at the cost of more cache misses and PM accesses.
Thus, most approaches bound probing distance to a fixed number, e.g., CCEH probes no more than four cachelines.
However, our evaluation (Section~\ref{sec:eval}) shows that linear probing alone is not enough in achieving high load factor.

Another important aspect of dynamic PM hashing is to ensure failure atomicity, particularly during segment split which is a three-step process:
(1) allocate a new segment in PM, (2) rehash records from the old segment to the new segment and (3) register the new segment in the directory and update local depth. 
Existing approaches such as CCEH only focused on step 3, side-stepping PM management issues surrounding steps 1--2.
If the system crashes during step 1 or 2, we need to guarantee the new segment is reclaimed upon restart to avoid \textit{permanent} memory leaks.
%CCEH mainly talks about how to achieve the failure-atomicity of the step (3) but ignore the step (1) and (2). 
%In Section~\ref{sec:detail}, we illustrate our patch for CCEH to guarantee the full correctness of the segment split. 
In Sections~\ref{sec:dash-eh} and \ref{subsec:impl}, we describe \dash's solution and a solution for existing approaches.

\section{Design Principles}
\label{sec:principles}
The aforementioned issues and performance characteristics of Optane DCPMM lead to the following design principles of \dash:
\begin{itemize}[leftmargin=*]\setlength\itemsep{0em}
\item \textbf{Avoid both Unnecessary PM Reads and Writes.}
Probing performance impacts not only search operations, but also all the other operations.
Therefore, in addition to reducing PM writes, \dash must also remove unnecessary PM reads to conserve bandwidth and alleviate the impact of high end-to-end read latency.

\item \textbf{Lightweight Concurrency.} 
\dash should scale well on multicore machines with persistence guarantees. % under various workloads. 
Given the limited bandwidth, concurrency control must be lightweight to not incur much overhead (i.e., avoid PM writes for search operations, such as read locks).
Ideally, it should also be relatively easy to implement.

\item \textbf{Full Functionality.}
\dash must not sacrifice or trade off important features that make a hash table useful in practice.
In particular, it needs to support near-instantaneous recovery and variable-length keys and achieve high space utilization.

\end{itemize}

\section{\dash for Extendible Hashing}
\label{sec:dash-eh}
Based on the principles in Section~\ref{sec:principles}, we describe \dash in the context of \dash-Extendible Hashing (\dash-EH). %, a scalable extendible hashing approach for PM.
We discuss how \dash applies to linear hashing in Section~\ref{sec:dash-lh}.

%\subsection{Overview}
%To solve this problem, DASH adds a stash area in each segment to allow overflow inserts when the target bucket is full.
%The stash area is shared within the segment by all ``normal'' buckets, and kept small so it does not become a major overhead.
%DASH also uses linear probing and displacement strategies (Section~\ref{subsec:space}) to improve load factor and prevent pre-mature splits.
%The result is an architecture with (1) reduced cache misses when accessing the directory and (2) improved load factor and load factor.

\subsection{Overview}
\begin{figure}[t]
\centering
\includegraphics[width=0.85\columnwidth]{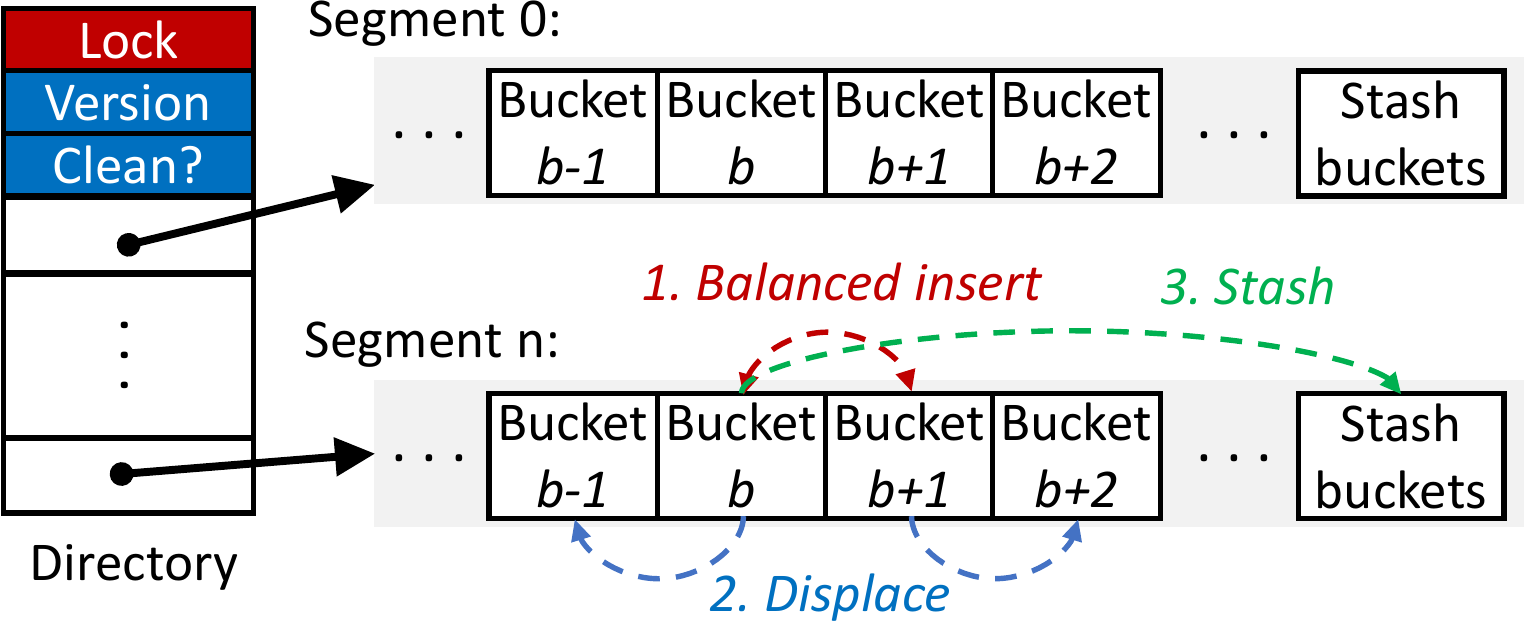}
%\vspace{-3mm}
\caption{Overall architecture of \dash-EH. 
%It uses a lock for concurrency control, a 1-byte version number and boolean marker for instant recovery support.
%Each directory entry points to a segment that consists of normal and stash buckets.
%The numbered steps show the bucket load balancing strategy for handling inserts (Section~\ref{subsec:space}).
}
\label{fig:arch} 
\end{figure}

Similar to prior approaches~\cite{CCEH,LarsonLockFreeLH}, \dash-EH uses segmentation. % as discussed in Section~\ref{subsec:pm-hash}.
As shown in Figure~\ref{fig:arch}, each directory entry points to a segment which consists of a fixed number of normal buckets and stash buckets for overflow records from normal buckets which did not have enough space for the inserts.
The lock, version number and clean marker are for concurrency control and recovery, which we describe later.

\begin{figure}[t]
\centering
\includegraphics[width=0.9\columnwidth]{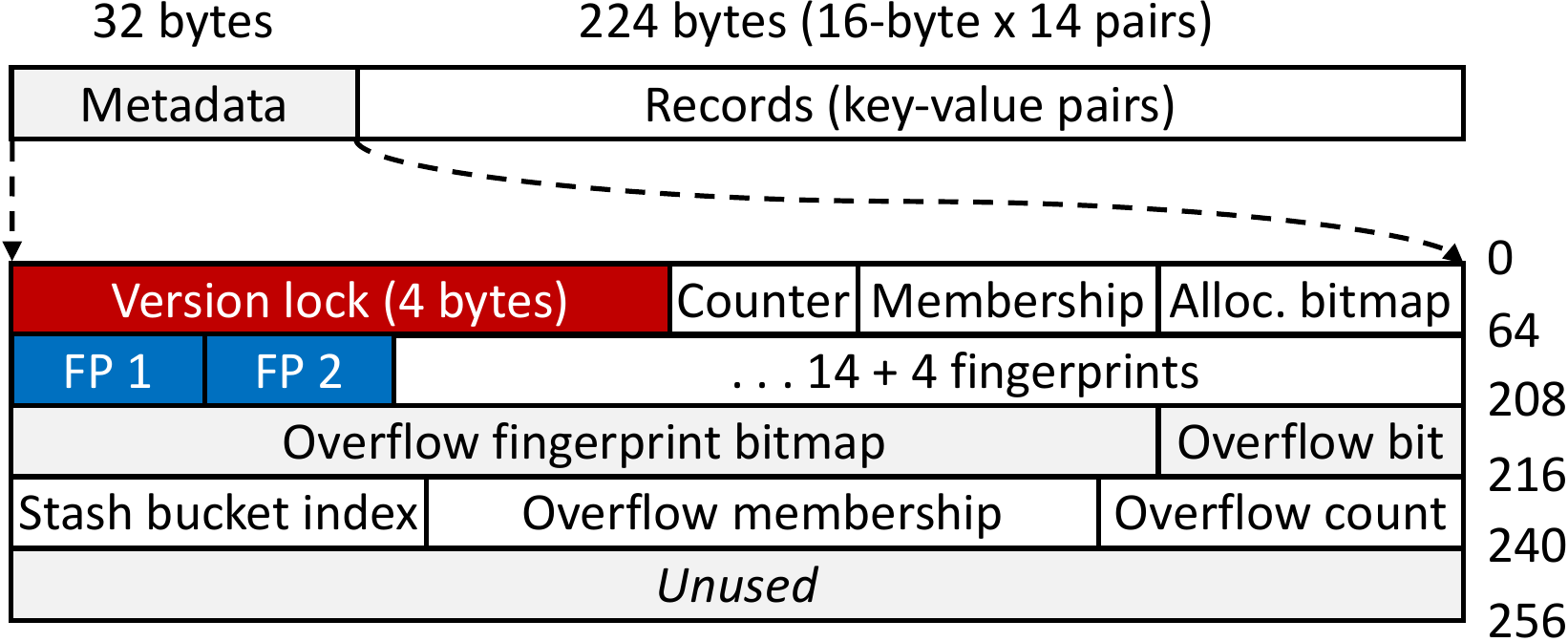}
%\vspace{-3mm}
\caption{\dash-EH bucket layout. The first 32 bytes are dedicated to 
metadata that optimizes probing and load factor, followed by records.
Normal and stash buckets share the same layout.}
\label{fig:bucket} 
\end{figure}

Figure~\ref{fig:bucket} shows the internals of a bucket.
We place the metadata used for bucket probing on the first 32 bytes, followed by multiple 16-byte record (key-value pair) slots.
The first 8 bytes in each slot store the key (or a pointer to it for keys longer than 8 bytes).
The remaining 8 bytes store the payload which is opaque to \dash; it can be an inlined value or a pointer, depending on the application's need.
The size of a bucket is adjustable.
In our current implementation it is set to 256-byte (block size of Optane DCPMM~\cite{UCSDMeasurement}) for better locality.
This allows us to store 14 records (16-byte each) per bucket.

The 32-byte metadata includes key data structures for \dash-EH to handle hash table operations and realize the design principles.
It starts with a 4-byte version lock for optimistic concurrency control (Section~\ref{subsec:cc}). 
A 4-bit \texttt{counter} records the number of records stored in the bucket.
The \texttt{allocation} bitmap reserves one bit per slot, to indicate whether the corresponding slot stores a valid record.
The \texttt{membership} bitmap is reserved for bucket load balancing which we describe later (Section~\ref{subsec:space}).
What follows are structures such as fingerprints and counters to accelerate probing and improve load factor.
Most unnecessary probings are avoided by scanning the fingerprints area.

\subsection{Fingerprinting}
\label{subsec:fp}
Bucket probing (i.e., search in one bucket) is a basic operation needed by all the operations supported by a hash table (search, insert and delete) to check for key existence. 
Searching a bucket typically requires a linear scan of the slots.
This can incur lots of cache misses and is a major source of PM reads, especially so for long keys stored as pointers. 
It is a major reason for hash tables on PM to exhibit low performance. %, due to the high end-to-end read latency of PM (Section~\ref{subsec:optane}).
%It is therefore imperative to reduce unnecessary reads to PM.
Moreover, such scans for negative search operations (i.e., when the target key does not exist) are completely unnecessary.

We employ fingerprinting~\cite{FPTree} to reduce unnecessary scans.
It was used by trees to reduce PM accesses with an amortized number of key loads of one.
We adopt it in hash tables to reduce cache misses and accelerate probing.
Fingerprints are one-byte hashes of keys for predicting whether a key possibly exists.
We use the least significant byte of the key's hash value. % as the fingerprint.
%For each key in the bucket, we take the least significant byte of its hash value as the fingerprint.
To probe for a key, the probing thread first checks whether any fingerprint matches the search key's fingerprint.
It then only accesses slots with matching fingerprints, skipping all the other slots.
If there is no match, the key is definitely not present in the bucket.
This process can be further accelerated with SIMD instructions~\cite{IntelManual}.

Fingerprinting particularly benefits negative search (where the search key does not exist) and uniqueness checks for inserts.
It also allows \dash to use larger buckets to tolerate more collisions and improve load factor, without incurring many cache misses: most unnecessary probes are avoided by fingerprints.
This design contrasts with many prior designs that trade load factor for performance by having small buckets of 1--2 cachelines~\cite{PFHT,CCEH,LevelHashing}.
% which can incur frequent bucket/segment splits that incur many PM reads and writes.
%Also, later accesses to the actual key and value is expected to have lower latency because after accessing fingerprints in the bucket, multiple key-value items could have been already fetched by the CPU as they reside in nearby cachelines. %cache buffer of Optane due to its larger access granularity(256 bytes).

As Figure~\ref{fig:bucket} shows, each bucket contains 14 slots, but 18 fingerprints (bits 64--208); 
14 are for slots in the bucket, and the other four represent keys placed in a stash bucket but were originally hashed into the current bucket.
They can allow early avoidance of access to stash buckets, saving PM bandwidth.
We describe details next as part of the bucket load balancing strategy that improves load factor.

\subsection{Bucket Load Balancing}
\label{subsec:space}
Segmentation reduces cache misses on the directory (by reducing its size).
However, as we describe in Sections~\ref{subsec:pm-hash} and~\ref{sec:eval}, this is at the cost of load factor: in a naive implementation the entire segment needs to be split if any bucket is full, yet other buckets in the segment may still have much free space.
We observe that the key reason is load imbalance caused by the (inflexible) way buckets are selected for inserting new records, i.e., a key is only mapped to a single bucket.
\dash uses a combination of techniques for new inserts to balance loads among buckets while limiting PM reads needed.
Algorithm~\ref{algo:insert} shows how the insert operation works in \dash-EH at a high level, with three key techniques described below.

\textbf{Balanced Insert.}
To insert a record whose key is hashed into bucket $b$ ($hash(key) = b$), \dash probes both bucket $b$ and $b+1$ and inserts the record into the bucket that is less full (Figure~\ref{fig:arch} step 1).
Lines 17--21 in Algorithm~\ref{algo:insert} show the idea; we discuss how a record is inserted into a bucket later.  %insertion of a record into a bucket and other details in the following sections.
The rationale behind is to improve load factor by amortizing the load of hot buckets while limiting PM accesses (at most two buckets).
Compared to balanced insert, linear probing allows a record to be inserted into bucket $b+n$ where $n > 1$ if buckets $b...b+n-1$ are full.
Probing multiple buckets may degrade performance by imposing more PM reads and cache misses.
It is also hard to tune the number of buckets to probe.

\textbf{Displacement.}
If both the target bucket $b$ and probing bucket $b+1$ are full, \dash-EH tries to displace (move) a record from bucket $b$ or $b+1$ to make room for the new record (Algorithm~\ref{algo:insert} lines 23--26).
With balanced insert, a record in bucket $n+1$ can be moved to $n+2$ if (1) it could be inserted to either bucket (i.e., $n+2$ is the probing bucket of the record being moved), \textit{and} (2) bucket $n+2$ has a free slot.
Thus, for a record with $hash(key)=b$ and both $b$ and $b+1$ are full, we first try to find a record in $b+1$ whose $hash(key)=b+1$ and move it to $b+2$.
If such a record does not exist, we repeat for bucket $b$ but move a record with $hash(key)=b-1$ (the target bucket).
In essence, displacement follows a similar strategy to balanced insert, but is for existing records.
%If a slot can be freed in bucket $b$ or $b+1$ in this process, we insert the new record there.

We use a per-bucket \texttt{membership} bitmap (Figure~\ref{fig:arch}) to accelerate displacement.
%Each bit indicates whether the bucket is the corresponding key's target or probing bucket.
If a bit is set, then the corresponding key was not originally hashed into this bucket; it was placed here because of balanced insert or displacement. 
When checking for bucket $b$ ($b+1$), a record whose membership bit is set (unset) can be displaced.
\dash then only needs to scan the bitmap to pick directly a record to move, without having to examine the actual keys.
This reduces unnecessary PM accesses, and is especially beneficial for variable length keys which are not inlined but represented by pointers.

\iffalse
\begin{algorithm}[t]
\input{dash-eh-search-helpers.tex}
\caption{\dash-EH bucket search.} \label{algo:bucket-search}
\end{algorithm}
\fi

\begin{algorithm}[t]
\begin{lstlisting}[language=python,
	mathescape,
	gobble=0,
	keywordstyle=\ttfamily\bfseries\color{blue},
	]
def dash_eh_insert(key, value):
  h = hash(key)
retry:
  # Obtain references and lock buckets
  [target_seg] = get_segment(h)
  [target_bucket, probing_bucket] = target_seg.bk(h)
  Lock target_bucket and probing_bucket

  # Verify the correctness of the segment reference
  [verify_seg] = get_segment(h)
  if verify_seg is not target_seg:
    Unlock and goto retry
  
  if key exists in either bucket or the stash: 
    Unlock and return Result::KeyExists

  if target_bucket or probing_bucket is not full:
    if target_bucket.count <= probing_bucket.count:
      target_bucket.insert(key, value, h)
    else
      probing_bucket.insert(key, value, h)
  else
    # Try displacement (possibly stashing)
    bucket = displace(target_bucket, probing_bucket)
    if bucket is not NULL:
      bucket.insert(key, value, h)
    elif stash_bucket.insert(key, value, h):
      target.overflow = true
      Set overflow fingerprint bitmap and fingerprint
    else  # Stashing failed, have to split
      split_segment(h)
      goto retry
  Unlock target_bucket and probing_bucket
  return Result::Inserted
\end{lstlisting}

\caption{\dash-EH insert algorithm with bucket load balancing.} \label{algo:insert}
\end{algorithm}

\textbf{Stashing.}
If the record cannot be inserted into bucket $b$ or $b+1$ after balanced insert and displacement, stashing will be the last resort before segment split has to happen.
In Figure~\ref{fig:arch}, a tunable number of stash buckets follow the normal buckets in each segment.
If a record cannot be inserted into its target bucket nor the probing bucket, we insert the record to a stash bucket; we call these records \textit{overflow records}.
Stash buckets use the same layout as that of normal buckets; probing of a stash bucket follows the same procedure as probing a normal bucket (see Section~\ref{subsec:fp}).
While stashing can be effective in improving load factor, it could incur non-trivial overhead: the more stash buckets are used, the more CPU cycles and PM reads will be needed to probe them. 
This is especially undesirable for negative search and uniqueness check in insert operations, since both need to probe all stash buckets, despite it may be completely unnecessary. 

%We observe the key to solving this problem is to conduct membership test at the target or probing bucket, and only probe stash buckets when necessary.
To solve this problem, we try to set up record metadata including fingerprints in a normal bucket and only refer actual record access to the stash bucket.
As Figure~\ref{fig:bucket} shows, four additional fingerprints per bucket are reserved for overflow records stored in stash buckets. 
A 4-bit \texttt{overflow fingerprint} bitmap records whether the corresponding fingerprint slot is occupied. 
Another \texttt{overflow bit} indicates whether the bucket has overflowed any record to a stash bucket. 
Algorithm~\ref{algo:insert} (lines 27--29) shows this process.
Similar to inserting records into a normal bucket, the overflow record's fingerprint also allows one more bucket probing, using the \texttt{overflow membership} bitmap to indicate whether the overflow fingerprints originally belong to this bucket. 
The 2-bit \texttt{stash bucket index} per overflow record indicates which one of the four stash buckets the record is inserted into for faster lookup. %to accelerate later operations. 
If the overflow fingerprint cannot be inserted into neither the target nor the probing bucket, we increment \texttt{overflow count} in the target bucket.
Once the counter becomes positive, a probing thread will have to check the stash area to ensure that a key does or does not exist.
Thus, it is desirable to reserve enough slots of overflow fingerprints in each bucket so that the overflow counter is rarely positive.
As Section~\ref{sec:eval} shows, using 2--4 stash buckets per segment can improve load factor to over 90\% without imposing significant overhead.

\subsection{Optimistic Concurrency}
\label{subsec:cc}
%Concurrency control plays an important role in the design of PM data structures: it must be lightweight enough to not cause excessive reads and writes in PM.
%Traditional locking based approaches, e.g., bucket-level locks, are pessimistic in nature and therefore requires taking read locks.
%Nevertheless, pessimistic locking is still one of the most widely used approaches in recent PM based hash tables~\cite{CCEH,LevelHashing} because of its simplicity and relatively good performance on emulated PM.
%However, pessimistic locking does not scale~\cite{MOCC,Silo,FOEDUS} in nature as even read operations---albeit free of conflicts and can coexist---cause writes to PM to set lock state. 
%Besides, the write performance of Optane DCPMM is severely limited and cannot scale because of its internal contention \cite{UCSDGuide}, which makes pessimistic locking more undesirable. 
%Some recent work~\cite{NVCHashmap} employs lock-free programming that uses atomic instructions directly to coordinate shared data access.
%Lock-free approaches can achieve high performance and typically avoids unnecessary writes during read operations, but it is notoriously hard to get right, more so in PM because of extra care has to be taken for safe persistence~\cite{PMwCAS}.

\dash employs optimistic locking, an optimistic flavor of bucket-level locking inspired by optimistic concurrency control~\cite{OCC,Silo}.
Insert operations will follow traditional bucket-level locking to lock the affected buckets.
Search operations are allowed to proceed without holding any locks (thus avoiding writes to PM) but need to verify the read record.
For this to work, in \dash the lock consists of (1) a single bit that serves the role of ``the lock'' and (2) a version number for detecting conflicts (not to be confused with the version number in Figure~\ref{fig:arch} for instant recovery).
As line 7 in Algorithm~\ref{algo:insert} shows, the inserting thread will acquire bucket-level locks for the target and probing buckets.
This is done by atomically setting the lock bit in each bucket by trying the \texttt{compare-and-swap} (\CAS{}) instruction~\cite{IntelManual} until success.
Then the thread enters the critical section and continues its operations.
After the insert is done, the thread releases the lock by (1) resetting the lock bit and (2) incrementing the version number by one, in one step using an atomic write.

To probe a bucket for a key, \dash first takes a snapshot of the lock word and checks whether the lock is being held by a concurrent writer (the lock bit is set).
If so, it waits until the lock is released and repeats.
Then it is allowed to read the bucket \textit{without} holding any lock.
Upon finishing its operations, the reader thread will read the lock word again to verify the version number did not change, and if so, it retries the entire operation as the record might not be valid as a concurrent write might have modified it.
%\textcolor{BrickRed}{
This lock-free read design requires segment deallocation (due to merge) happen only after no readers are (and will be) using the segment. 
We use epoch-based reclamation~\cite{epoch} to achieve this without incurring much overhead.
%}

\dash does not use segment-level locks, saving PM access in the segment level.
As a result, structural modification operations (SMOs, such as segment split) need to lock all the buckets in each segment.
Directory doubling/halving is handled similarly: the directory lock is only held when the directory is being doubled or halved.
For other operations on the directory (e.g., updating a directory entry to point to a new segment), no lock is taken.
Instead, they are treated as search operations without taking the directory lock. % and proceed as long as the directory lock is free, and must validate that the directory version did not change later.
This is safe because we guarantee isolation in the segment level: an inserting thread must first acquire locks to protect the affected buckets.
``Real'' probings (search/insert) proceed without reading the directory lock but again need to verify that they entered the right segment by re-reading the directory to test whether these two read results match; if not, the thread aborts and retries the entire operation. 

%Optimistic locking avoids unnecessary writes to PM as for search operations, no lock is ever taken, saving much precious PM write bandwidth.
%As we show in Section~\ref{sec:eval}, the result is up to 2.4$\times$ faster search throughput and near-linear scalability.

\subsection{Support for Variable-Length Keys}
%\dash stores fixed-length keys inline.
%To support variable-length keys, 
\dash stores pointers to variable-length keys, which is a common approach~\cite{FPTree,BzTree,CCEH,LevelHashing}.
A knob is provided to switch between the inline (fixed-length keys up to 8 bytes) and pointer modes.
%A major drawback of this design is that checking the bitmap may incur more cache misses (thus more PM reads), in addition to the cache misses caused by accessing the key itself. 
%The high end-to-end PM read latency further worsens the situation.
%Some prior work avoids using bitmaps and instead (1) ``steal'' a bit (e.g., the MSB) on the key field to indicate whether the value stored is an actual inlined key or a pointer, or (2) provides a knob to switch between pure pointer or inline modes~\cite{FPTree}.
%The former leverages the fact that modern x86 microarchitecture only uses 48 bits for addressing, but effectively limits the key space for inlined keys to 63 bits.
%The latter sacrifices flexibility and may penalize performance for small ($\le$8 bytes) key sizes as every key access will likely incur a cache miss.
%DASH leverages fingerprints and uses bitmap to avoid limiting key space and alleviates the performance penalty of derefenencing pointers.
Though dereferencing pointers may incur extra overhead, fingerprinting largely alleviates this problem.
For negative search where the target key does not exist, no fingerprint will match and so key probing will not happen at all.
For positive search, as we have discussed in Section~\ref{subsec:fp}, the amortized number of key load (therefore the number cache misses caused by following the key pointer) is one~\cite{FPTree}.

\subsection{Record Operations} 
Now we present details on how \dash-EH performs insert, search and delete operations on PM with persistence guarantees.
 
\begin{algorithm}[t]
\begin{lstlisting}[language=python,
	mathescape,
	gobble=0,
	keywordstyle=\ttfamily\bfseries\color{blue},
	]
def bucket::insert(key, value, h):
  slot = slots[slot_id = get_free_slot()]
  slot.assign(key, value)
  CLWB+FENCE(slot)  # Persist the record first

  fingerprints[slot_id] = LSB_byte(h)
  # Since following metadata are in the same word,
  # their updates are done in one store operation
  alloc_bitmap.set(slot_id)
  ++counter 
  if bucket is probing bucket:
    membership.set(slot_id)
  
  # Persist all metadata updates in one flush 
  # (same cacheline, no reordering on x86)
  CLWB+FENCE(alloc_bitmap, membership, counter, FP)

def displace(target = b, prob = b+1):
  # Try to move a record from bucket b+1 to b+2 
  slot_id = prob.get_unset_LSB(membership)
  if slot_id is not Invalid and (b+2) is not full:
    (prob+1).insert(prob.slots[slot_id])
    # Mark deletion for the moved record, decrement 
    # the counter (done in one store operation)
    prob.alloc_bitmap &= $\sim$(1 << slot_id)
    prob.counter--
    CLWB+FENCE(prob.alloc_bitmap, counter)
    return prob
  else
    # Try to move a record from bucket b to b-1
    slot_id = target.get_set_LSB(membership)
    if slot_id is not Invalid and (b-1) is not full:
      (target-1).insert(target.slots[slot_id])
      target.alloc_bitmap &= $\sim$(1 << slot_id)
      target.membership &= $\sim$(1 << slot_id)
      target.counter--
      CLWB+FENCE(target.alloc_bitmap, counter)
      return target
    else
      return NULL
\end{lstlisting}

\caption{Bucket insert and displacement in \dash-EH.} \label{algo:insert-helpers}
\end{algorithm}

\textbf{Insert.}
Section~\ref{subsec:space} presented the high-level steps for insert; here we focus on the bucket-level. % operations.
%After choosing a bucket to insert the record, 
As the \texttt{bucket::insert} function in Algorithm~\ref{algo:insert-helpers} shows, we first write and persist the new record (lines 3-4), and then set up the metadata (fingerprint, allocation bitmap, counter and membership, lines 6--12).
Note that the allocation bitmap, membership bitmap and counter are in one word; they are updated in one atomic write.
The \CLWB{} and fence are then issued (line 16) to persist all the metadata. % in one flush.
Once the corresponding bit in the bitmap is set, the record is visible to other threads.
If a crash happens before the bitmap is persisted, the new record is regarded as invalid; otherwise, the record is successfully inserted. 
This allows us to avoid expensive logging while maintaining consistency.

Displacing a record needs two steps: (1) inserting it into the new bucket and (2) deleting it from the original bucket.
As the \texttt{displace} function in Algorithm~\ref{algo:insert-helpers} shows, 
step 2 is done by resetting the corresponding bit in the allocation bitmap without moving data.
In case a crash happens before step 2 finishes, a record will appear in both buckets.
This necessitates a duplicate detection mechanism upon recovery, which is amortized over runtime (see Section~\ref{subsec:recovery}).
%Level hashing \cite{LevelHashing}, which also has the record duplicate problem because of the cuckoo displacement, solves this problem by sacrificing the performance of update operation and delete operation. 
%Specifically, the update operation and delete operation needs to probe all possible buckets to detect the duplicate and make the hash table restore to a consistent state.
%Different from it, the DASH hash table conducts the lightweight duplicate detection during the lazy recovery step, which will be specifically illustrated later.

If the insert has to happen in a stash bucket, we set the overflow metadata in the normal bucket. 
This cannot be done atomically with 8-byte writes and may need a (complex) protocol for crash consistency.
We note that the overflow metadata is an optimization and does not influence correctness: records can still be found correctly even without it.
So we do not explicitly persist it and rely on the lazy recovery mechanism to build it up gradually (described later). 

\begin{algorithm}[t]
\begin{lstlisting}[language=python,
	mathescape,
	gobble=0,
	keywordstyle=\ttfamily\bfseries\color{blue},
	]
def dash_eh_search(key):
retry:
  # Get the segment and buckets
  [target_seg] = get_segment(h)
  [target_bucket, probing_bucket] = target_seg.bk(h)
  vt = target_bucket.version_lock
  vp = probing_bucket.version_lock
  
  # Verify the correctness of the segment reference
  [verify_seg] = get_segment(h)
  if verify_seg is not target_seg:
    goto retry
  
  if is_lock_set(vt) or is_lock_set(vp):
    goto retry
  
  result = target_bucket.search(key)
  if vt is not target_bucket.version_lock:
    goto retry
  if result is not NULL:
    return result
  
  result = probing_bucket.search(key)
  if vp is not probing_bucket.version_lock:
    goto retry
  if result is not NULL:
    return result
  
  # Determine whether to search in the stash buckets
  # Note that version lock check is omitted below
  if probing bucket.overflow_count is zero:
    if key matches overflow fingerprints:
      search corresponding stash buckets and return
    else
      return NULL
  else 
    Search by scanning the stash buckets and return
  return NULL
\end{lstlisting}

\caption{\dash-EH search algorithm.} \label{algo:search}
\end{algorithm}

\textbf{Search.}
With balanced insert and displacement, a record could be inserted into its target bucket $b$ where $b=hash(key)$ or its probing bucket $b+1$.
A search operation then has to check both if the record is not found in $b$.
%Although these strategies may improve load factor at the cost of more probing (must probe both bucket $b$ and $b+1$), the overhead is offset by the use of the membership bitmap and limited number of probings.
As Algorithm~\ref{algo:search} shows, % how a search request is handled.
%to search for a key, 
the probing thread first reads the directory to obtain a reference to the corresponding segment and buckets (lines 4--5). %determined by the hash value (lines 4--5).
It then takes a snapshot of the version number of both buckets (lines 6--7) for verification later.
%Following the optimistic approach (Section~\ref{subsec:cc}), 
We verify at line 10 that the segment did not change (i.e., the directory entry still points to it) and retry if needed.
Once segment check passed, we check whether the target/probing buckets are being modified (i.e., locked) at lines 14--15.
If not, we continue to search the target and probing buckets (lines 17--27) using the \texttt{bucket::search} function (not shown). %in Algorithm~\ref{algo:bucket-search}.
Note that we need to verify the lock version did not change after \texttt{bucket::search} returns (lines 18 and 24).

If neither bucket contains the record, it might be stored in a stash bucket (lines 31--37).
If \texttt{overflow\_count} $>0$, then we search the stash buckets as the overflow fingerprint area does not have enough space for all overflow records from the bucket.
Otherwise, stash access is only needed if there is a matching fingerprint (lines 31--35).

\textbf{Delete.}
To delete a record from a normal bucket, we reset the corresponding bit in the allocation bitmap, decrement the counter and persist these changes.
Then the slot becomes available for future reuse.
%A segment merge operation will be triggered if the load factor drops below a threshold.
To delete a record from a stash bucket, in addition to clearing the bit in the allocation bitmap, we also clear the overflow fingerprint in the target bucket which this record overflowed from if it exists; otherwise we only decrement the target bucket's overflow counter.
%Then the correctness of the later access operation is still guaranteed.

\subsection{Structural Modification Operations}
\label{subsec:smo}
%The structure modifications such as segment split/merge incurs a large number of memory operations. 
%The following only talks how to achieve the crash consistency of the segment split operation because the segment merge is just the reverse of the segment split. 
When a thread has exhausted all the options to insert a record into a bucket, it triggers a segment split and possibly expansion of the directory.
Conversely, when the load factor drops below a threshold, segments can be merged to save space.
At a high level, three steps are needed to split a segment $S$:
(1) allocate a new segment $N$, (2) rehash keys in $S$ and redistribute records in $S$ and $N$, and (3) attach $N$ to the directory and set the local depth of $N$ and $S$. 
These operations cause the structure of the hash table to change and must be made crash consistency on PM while maintaining high performance.
%We present a lightweight approach to achieve this goal and enable truly instant recovery.
%Logging is the simplest way to guarantee the crash consistency of the segment split but incurs non-negligible overhead. 
%Thus we introduce a lightweight method to make the DASH hash table recoverable in the middle of the structure modification with minor logging. 

\begin{figure}[t]
\centering
\includegraphics[width=0.9\columnwidth]{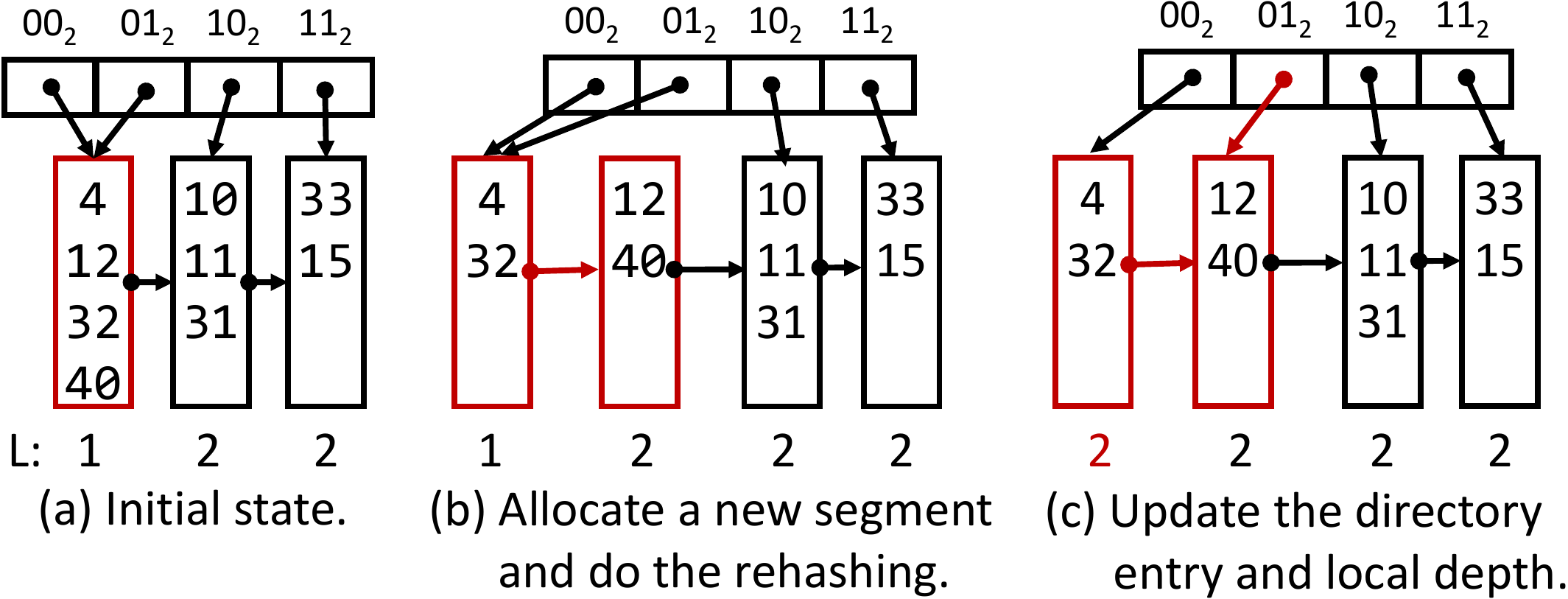}
%\vspace{-3mm}
\caption{Segment split in \dash-EH; the global depth is 2.
}
\label{fig:split} 
\end{figure}

%To make steps 1--2 crash consistent, 
For crash consistency, \dash-EH chains all segments using side links to the right neighbor. %adds in each segment a side link to its right neighbor segment so that all segments are linked together.
Each segment has a \texttt{state} variable to indicate whether the segment is in an SMO and whether it is the one being split or the new segment.
An initial value of zero indicates the segment is not part of an SMO.
Figure~\ref{fig:split} shows an example. % segment split in \dash-EH.
Note that as shown by the figure, \dash-EH uses the most significant bits (MSBs) of hash values to address and organize segments and buckets (i.e., the directory is indexed by the MSBs of hash values), similar to other recent work~\cite{CCEH}.
This is different from traditional extendible hashing described in Section~\ref{sec:bg-dyn-hash} that uses LSBs of hash values to address buckets.
Using LSBs was the choice in the disk era to reduce I/O: the directory can be doubled by simply copying the original directory and appending it to the directory file.
On PM, such advantage is marginalized as to double a directory, one needs to allocate and persist a double-sized directory in PM anyway to keep the directory in a contiguous address space.
%This and flush the whole directory no matter LSB or MSB is used, which makes the advantage of LSB marginalized.
%However, with segmentation the directory entries for a segment are collocated together on MSB-based extendible hashing so that the directory doubling process needs to update all entries for the double-sized directory. 
%Moreover, MSB organization is important for the recovery which we will show later so that \dash-EH uses MSB to organize the segments. 
Using MSBs also allows directory entries pointing to the same segment to be co-located, reducing cacheline flushes during splits~\cite{CCEH}. 
%We therefore adopt the MSB-based approach in \dash.
%MSB organization is beneficial for saving the logging overhead which we will show later so that \dash-EH uses MSB to organize the segments.

To split a segment $S$, we first mark its \texttt{state} as \texttt{SPLITTING} and allocate a new segment $N$ whose address is stored in the side link of $S$.
%After setting the state variable of the splitting segment as \texttt{SPLITTING}, the new segment is allocated and stored in the side link of the splitting segment. 
$N$ is then initialized to carry $S$'s side link as its own.
Its local depth is set to the local depth of $S$ plus one.
Then, we change $N$'s \texttt{state} to \texttt{NEW} to indicate this new segment is part of a split SMO for recovery purposes (see Section~\ref{subsec:recovery}).
We rely on PM programming libraries (PMDK~\cite{PMDK}) to atomically allocate \textit{and} initialize the new segment; in case of a crash, the allocated PM block is guaranteed to be either owned by \dash or the allocator and will not be permanently leaked.
After initialization, we finish up step 2 by redistributing records between $N$ and $S$.
Records moved from $S$ to $N$ are deleted in $S$ after they are inserted into $N$.
Note that the rehashing/redistributing process does not need to be done atomically: if a crash happens in the middle of rehashing, upon (lazy) recovery we redo the rehashing process with uniqueness check to avoid repeating work for records that were already inserted into $N$ before the crash; we describe more details later in Section~\ref{subsec:recovery}.
%Our lazy recovery introduced later could handle the interrupted rehashing and help finish the entire SMO. 
Figure~\ref{fig:split}(b) shows the state of the hash table after step 2. 
Then the directory entry for $N$ and the local depth of $S$ are updated as shown in Figure~\ref{fig:split}(c). %, which is doen atomically using a single store instruction.
Similarly, these updates are conducted using an atomic PMDK transaction which may use any approach such as lightweight logging.
Many other systems avoid the use of logging to maintain high performance, largely because of the frequent pre-mature splits.
But split is much rarer in \dash thanks to bucket load balancing that gives high load factor (Section~\ref{subsec:space}); this allows \dash-EH to employ logging-based PMDK transactions that abstracts away many details and eases implementation.

\subsection{Instant Recovery}
\label{subsec:recovery}
%Different from many prior proposals that require complex work upon recovery (therefore trading off the potential of instant recovery), such as linearly scanning the directory, \dash provides truly instant recovery.
\dash provides truly instant recovery by requiring a constant amount of work (reading and possibly writing a one-byte counter), before the system is ready to accept user requests.
%We add a global version number and a \texttt{clean} marker next to the directory lock in Figure~\ref{fig:arch}, as well as a per-segment version number.
We add a global version number $V$ and a \texttt{clean} marker shown in Figure~\ref{fig:arch}, and a per-segment version number.
\texttt{clean} is a boolean that denotes whether the system was shutdown cleanly; $V$ tells whether recovery (during runtime) is needed.
Upon a clean shutdown, \texttt{clean} is set to true and persisted.
Upon restart, if \texttt{clean} is true, we set \texttt{clean} to false and start to handle requests.
Otherwise, we increment $V$ by one and start to handle requests.
For both clean shutdown and crash cases, ``recovery'' only involves reading \texttt{clean} and possibly bumping $V$.
The actual recovery work is amortized over segment accesses.

To access a segment, the accessing thread first checks whether the segment version matches $V$.
If not, the thread (1) recovers the segment to a consistent state before doing its original operation (e.g., insert or search), and (2) sets the segment's version number to $V$ so that future accesses can skip the recovery pass.
With such lazy recovery approach, a segment is not recovered until it is accessed.
Multiple threads may access a segment that needs to be recovered.
We employ a segment-level lock that is only for recovery purpose, but a thread only tries to acquire the lock if it sees the segment's version number does not match $V$.
% to allow only one thread to recover the segment.
Our current implementation uses one-byte version numbers.
In case the version number wraps around and recovery is needed, we reset $V$ to zero and set the version number of each segment to one.
Since crash and repeated crashes are rare events, such wrap-around cases should be very rare. 

%After the segment is correctly recovered, the thread will clear the most significant bit of the corresponding directory entry and continue executing the normal operation. 
%As the \dash hash table serves more and more operations, the hash table could be gradually recovered and restored to the normal throughput. 

Recovering a segment needs four steps: (1) clear bucket locks, (2) remove duplicate records, (3) rebuild overflow metadata, and (4) continue the ongoing SMO.
Some locks might be in the locked state upon crash, so every lock in each bucket needs to be reset. % to avoid the deadlock. 
Duplicate records are detected by checking the fingerprints in neighboring buckets.
This is lightweight since the real key comparison is only needed if the fingerprints match. 
Overflow metadata in normal buckets also needs to be cleared and rebuilt based on the records in stash buckets as we do not guarantee their crash consistency for performance reasons. 
Finally, if a segment is in the \texttt{SPLITTING} state, the accessing thread will follow the segment's side link to test whether the neighbor segment is in the \texttt{NEW} state.
%Since the logging only guarantee the consistency of step (3) of the segment split, some segments are still in a structure modification if the system is crashed during step (1) and (2). 
%Therefore, by checking the \texttt{state} variable, the thread could know whether the segment is in a structure modification and if so, it will help finish this SMO operation. 
If so, we restart the rehashing-redistribution process and finish the split. % as described in Section~\ref{subsec:smo}.
Otherwise, we reset the \texttt{state} variable which in effect rolls back the split.

\section{\dash for Linear Hashing}
\label{sec:dash-lh}
We present \dash-LH, \dash-enabled linear hashing that uses the building blocks discussed previously (balanced insert, displacement, fingerprinting and optimistic concurrency).
We do not repeat them here and focus on the design decisions specific to linear hashing.

\subsection{Overview}
Figure~\ref{fig:linear} (focus on segments 0-3 for now) shows the overall structure of \dash-LH. 
Similar to \dash-EH, \dash-LH also uses segmentation and splits at the segment level.
%Instead of pointing to the bucket, the directory entry actually points to the segment, which consists of a group of ``normal'' buckets and a stash area. 
%The split unit of DASH-LH is the segment rather than the bucket. 
However, we follow the linear hashing approach to always split the segment pointed to by the \texttt{Next} pointer, which is advanced after the segment is split.
Since the segment to be split is not necessarily a full segment, it needs to be able to accommodate overflow records, e.g., using linked lists.
%Thus the previous linear hashing allows the bucket to be overflowed by storing the overflowed items in the linked list. 
However, linked list traversal would incur many cache misses, which is a huge penalty for PM hash tables.
Instead, we leverage the stashing design in \dash and use an adjustable number of stash buckets.
%We observe that the DASH segment could handle the overflowed items from the segment with minor modifications and do not violate the design principles of DASH. 
%The solution is to allow the stash area to be variable-sized. 
In addition to a fixed number of stash buckets (e.g., 2 stash buckets) in each segment, we store a linked list of stash buckets. 
A segment split is triggered whenever a stash bucket is allocated to accommodate overflow records.
This contrasts with classic linear hashing which splits a bucket at a time which is vulnerable to long overflow chains under high insertion rate. 
\dash-LH uses larger split unit (segment) and chaining unit (stash bucket rather than individual records), reducing chain length (therefore pointer chasing and cache misses).
The overflow metadata and fingerprints further helps alleviate the performance penalty brought by the need to search stash buckets.
Overall, as we show in Section~\ref{sec:eval}, \dash-LH can also achieve near-linear scalability on realistic workloads.
%Moreover, even if one segment has the chaining stash buckets, the access performance on this segment is nearly unaffected because of the overflow metadata. 

\subsection{Hybrid Expansion}
Similar to \dash-EH, it is also important to reduce directory size for better cache locality. %, although each directory entry points to a segment, the directory can still spill out of the CPU cache if the hash table is very large. 
Some designs use double expansion~\cite{DoubleExpansion} %previous proposals typically employ fixed expansion \cite{LarsonLinearHashing} or double expansion \cite{DoubleExpansion}. 
%The former works similarly to segmentation: each directory entry points to a large fixed-length segment where every bucket uses chaining to handle the hash collisions and the split unit is still the bucket. 
%Double expansion uses the segment of exponentially increasing size so that the allocation of the new segment doubles the number of buckets in the hash table. 
which increases segment size exponentially: allocating a new segment doubles the number of buckets in the hash table. 
For example, the second segment allocated would be $1\times$ the size of the first segment, and the third segment would be $2\times$ larger than the first segment, and so on.
The benefit is that directory size can become very small and be often fit even in the L1 cache.
However, it also makes load factor drop by half whenever a new segment is allocated.

\begin{figure}[t]
\centering
\includegraphics[width=0.8\columnwidth]{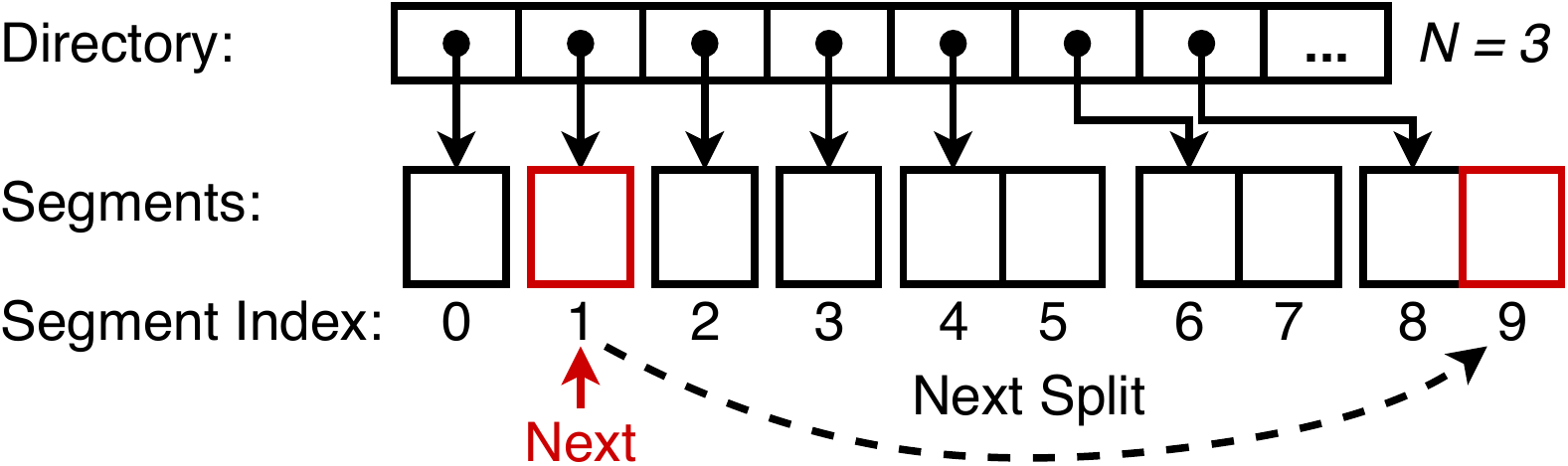}
%\vspace{-3mm}
\caption{Overview of \dash-LH. Segments are organized in arrays.}
\label{fig:linear} 
\end{figure}

%To achieve the benefits of double expansion while reducing space waste, %we propos a hybrid strategy that combines double expansion and segmentation.
To reduce space waste, 
we postpone double expansion and expand the hash table by several fixed-size segments first, before triggering double expansion.
%The hybrid expansion conducts the exponential-size expansion for the overall trend, but before allocating the larger segment (for DASH-LH, allocate a larger segment array), it will first conduct several fixed expansions. 
We call the number of such fixed expansions the \texttt{stride}.
Figure~\ref{fig:linear} shows an example (stride = 4). %of \dash-LH using the hybrid expansion with a stride of four. 
A directory entry can point to an array of segments; the first four entries point to one-segment arrays, the next four entries point to two-segment arrays, and so on. 
With a larger stride, the allocation of larger segment arrays will have less impact on load factor.
The result is very small directory size that is typically L1-resident.
Using 16KB segments, the first segment array will include 64 segments, with a stride of four, we can index TB-level data with a directory less than 1KB. 
%Compared to \dash-EH, the directory in \dash-LH is a static structure with the small memory preallocation (e.g., 1MB). 

\subsection{Concurrency}
\label{sec:lh-concurrency}
Since linear hashing expands in one direction, splits are essentially serialized by locking the \texttt{Next} pointer.
%  and one segment split begins after the last segment split is done, split can easily become a bottleneck using locking. 
%As shown in Figure~\ref{fig:linear}, if one segment is full and will overflow, the segment to be split is segment $1$ and its records will be redistributed to the new segment $9$. 
%Then the \texttt{Next} pointer advances one segment so that next segment to be split is segment $2$ and will redistribute the records to the new segment $10$. 
% encounter that the segment is overflowed during their insertions and thus 
%For instance, assume two threads ($T1$ and $T2$) both request to expand the hash table by splitting a segment. %, their split operations need to be handled one by one synchronized by the lock. 
%As shown in Figure~\ref{fig:linear}, if $T1$ gets the lock and $T2$ needs to wait, then thread 1 will split the segment 1, redistribute the records to the new segment 9, and then advance the \texttt{Next} pointer. 
%After the segment split of thread 1 is done, thread 2 could get the lock and start to split the segment 2. 
To shorten the length of the critical section, we adopt the expansion strategy proposed by LHlf~\cite{LarsonLockFreeLH} where the expansion only atomically advances \texttt{Next} without actually splitting the segment.
Then any thread that accesses a segment that should be split (denoted in the segment metadata area) will first conduct the actual split operation. 
As a result, multiple segments splits can execute in parallel by multiple threads. % if they hit different uninitialized segments.
%In our implementation, we use one bit in the segment metadata area to indicate wether it is initialized. 
%If not, the thread will grab all locks of the segment and conducts the segment split for it.  
Before advancing the \texttt{Next} pointer, the accessing thread first probes the directory entry for the new segment to test whether the corresponding segment array is allocated. 
If not, it allocates the segment array and stores it in the directory entry. 
%Multiple threads may simultaneously observe that PM allocation is needed, however, only one thread proceeds to do the memory allocation and other threads are blocked.
The performance of PM allocator therefore may impact overall performance, as we show in Section~\ref{sec:eval}. 
%Whether the synchronization on the memory allocation becomes the scalability bottleneck or not depends on the speed of the memory allocation. 
%Our evaluation result indicates that the memory allocation of PM allocator is extremely slow so that it limits the scalability of \dash-LH. 

%In addition to the \texttt{Next} pointer, 
\dash-LH uses a variable \texttt{N} to compute the number of buckets of the base table. 
After each round of the split, \texttt{Next} is reset to zero and \texttt{N} is incremented to denote that the number of buckets is doubled. 
For consistency guarantees, we store \texttt{N} (32-bit) and \texttt{Next} (32-bit) in a 64-bit word which can be updated atomically.

\section{Evaluation}
\label{sec:eval}
This section evaluates \dash and compares it with two other state-of-the-art PM hash tables, CCEH~\cite{CCEH} and level hashing~\cite{LevelHashing}. 
Specifically, through experiments we confirm the following:
\begin{itemize}[leftmargin=*]\setlength\itemsep{0em}
\item \dash-enabled hash tables (\dash-EH and \dash-LH) scale well on multicore servers with real Optane DCPMM;
\item The bucket load balancing techniques allow \dash to achieve high load factor while maintaining high performance;
\item \dash provides instant recovery with a minimal, constant amount of work needed upon restart, reducing service downtime.
\end{itemize}

%Next, we first describe important details about our implementation of \dash and other hash tables under comparison.

\subsection{Implementation}
\label{subsec:impl}
We implemented \dash-EH/LH using PMDK~\cite{PMDK}, which provides primitives for crash-safe PM management and synchronization.
%  to support crash-safe PM allocation and transactional PM management, as well as synchronization primitives that are crash consistent (e.g., locks will be released upon recovery).
These primitives are essential for building PM data structures, but also introduce overhead.
%We used the bundled allocator in PMDK to allocate memory from the persistent memory pool,
%These read-write locks are automatically unlocked after the system crash so that the persistent data structures could correctly recover.
For example, PMDK allocator exhibits scalability problems and is much slower than DRAM allocators~\cite{PiBench}.
Such overheads are ignored in previous emulation-based work, but are not negligible in reality.
We take them into account in our evaluation.
%To support real persistent memory operations, we extended each hash table with PMDK.
The other hash tables under comparison (CCEH~\cite{CCEH} and level hashing~\cite{LevelHashing}) were both proposed based on DRAM emulation.
%In addition to \dash-EH and \dash-LH, 
We ported them to run on Optane DCPMM using their original code\footnote{Code downloaded from \url{https://github.com/DICL/CCEH} and \url{https://github.com/Pfzuo/Level-Hashing}.} and PMDK. 
Like previous work~\cite{CCEH}, we optimize level hashing by co-locating all the locks in a small and continuous memory region (lock striping)~\cite{Mauricebook} to reduce cache misses. 
%All of our implementations, including \dash and modified versions of CCEH and level hashing are available at \url{https://github.com/baotonglu/dash}.
%we also adopted two state-of-the-art PM-optimized hash tables (i.e., CCEH and level hashing) to support real persistent memory. 
Now we summarize the key implementation issues and our solutions.

\textbf{Crash Consistency.}
\dash uses PMDK transactions for segment splits.
This frees \dash from handling low-level details while guaranteeing safe and atomic allocations.
We noticed a consistency issue in CCEH code where a power failure during segment split could leak PM.
%the hash table in an inconsistent state, since the segment splitting of step (1) and (2)
%since the segment splitting in CCEH requires multiple steps, and these steps need transaction or logging to guarantee atomicity.
We fixed this problem using PMDK transaction. % to ensure the atomicity of segment splitting.
%We believe this is a non-intrusive and efficient way to address this problem.
We also adapted CCEH and level hashing to use PMDK reader-writer locks that are automatically unlocked upon recovery.

%PMDK provides the persistent pointer that contains two fields: the base address and the offset, to easily support multiple pools in one application. However, atomically update the persistent pointer (16 bytes) is unavailable in current machines, thus we store one hash table in a single memory pool where all objects have the same base address. Then the pointers used in the data structure could only store the offset field. 
\textbf{Persistent Pointers.} 
Both CCEH and level hashing assume standard 8-byte pointers based on DRAM emulation, while some systems use 16-byte pointers for PM~\cite{PMDK,FOEDUS}.
Long pointers break the memory layout and make atomic instructions hard to use.
To use 8-byte pointers on PM, we extended PMDK to ensure that PM is mapped onto the same virtual address range across different runs (using \texttt{MAP\_FIXED\_NOREPLACE}\footnote{We also had to replace \texttt{MAP\_SHARED\_VALIDATE} with \texttt{MAP\_SHARED} for \texttt{MAP\_FIXED\_NOREPLACE} to work, detailed in our code repository.} in \texttt{mmap} and setting \texttt{mmap\_min\_addr} in the kernel).
All hash tables experimented here use this approach.

\textbf{Garbage Collection.}
%\BTD{
%The optimistic concurrency design in \dash requires the use of some mechanism that defers actual PM deallocation (i.e., segment) to happen only when no readers are still using it.
%}
%\textcolor{BrickRed}{
We implemented a general-purpose epoch-based PM reclamation mechanism for \dash.
We also observed that the open-sourced implementation of CCEH allows threads to access the directory without acquiring any locks, which may allow access to freed memory (due to directory doubling or halving).
We fixed this problem with the same epoch-based reclamation approach.
%}
%Since the memory reclamation in dynamic hashing is not frequent, to reduce the overhead, 
%we allow the threads to enroll in and exit from the epoch for every 1000 operations at the application level.

\subsection{Experimental Setup}
We run experiments on a server with a Intel Xeon Gold 6252 CPU clocked at 2.1GHz,
768GB of Optane DCPMM (6$\times$128GB DIMMs on all six channels) in AppDirect mode, and 192GB of DRAM (6$\times$32GB DIMMs).
The CPU has 24 cores (48 hyperthreads) and 35.75MB of L3 cache.
The server runs Arch Linux with kernel 5.5.3 and PMDK 1.7.
All the code is compiled using GCC 9.2 with all optimization enabled.
Threads are pinned to physical cores.

\textbf{Parameters.}
For fair comparison, we set CCEH and level hashing to use the same parameters as in their original papers~\cite{LevelHashing,CCEH}.
Our own tests showed these parameters gave the best performance and load factor overall.
Level hashing uses 128-byte (two cachelines) buckets.
CCEH uses 16KB segments and 64-byte (one cacheline) buckets, with a probing length of four.
\dash-EH and \dash-LH use 256-byte (four cachelines) buckets and 16KB segments.
Each segment has two stash buckets, making it enough to have four overflow fingerprint slots per bucket so that the overflow counter is rarely positive. 
\dash-LH uses hybrid expansion with a stride of eight and its first segment array includes 64 segments. 

\textbf{Benchmarks.}
We stress test each hash table using microbenchmarks.
%The benchmark will proceed by starting the timer and issuing the corresponding operations.
%For all operations, unless explicitly mentioned, are preceded by a single-thread 10M insert-only warm-up,
%and will be executed for 190M rounds for each of $SEARCH$, $INSERT$, and $DELETE$. 
For search operations, we run positive search and negative search: the latter probes specifically non-existent keys.
Unless otherwise specified, for all runs we preload the hash table with 10 million records, then execute 190 million inserts (as an insert-only benchmark), 190 million positive search/negative search/delete operations back-to-back on the 200-million-record hash table.
%A carefully chosen hash function should output uniformly distributed hash values to mitigate the impact of data skew. 
For all hash indexes, We use GCC's \texttt{std::\_Hash\_bytes} (based on Murmur hash~\cite{MurmurHash}) as the hash function, which is known to be fast and provides high-quality hashes.
Similar to other work~\cite{CCEH,LevelHashing}, we use uniformly distributed random keys in our workloads.
We also tested skewed workloads under the Zipfian distribution (with varying skewness) and found all operations achieved better performance benefitting from the higher cache hit ratios on hot keys, and contention is rare because the hash values are largely uniformly distributed.
Due to space limitation, we omit the detailed results over skewed workloads.
%}
For fixed-length key experiments, both keys and values are 8-byte integers; for variable-length key experiments, we use (pointers to) 16-byte keys and 8-byte values. 
The variable-length keys are pre-generated by the benchmark before testing.  

%The evaluation on skewed workload is also conducted but now shown in the paper because the result is similar to the uniform workload. 

%To study the performance of $SEARCH$, $INSERT$, and $DELETE$ operations, 
%we first pre-generate all the keys to avoid runtime random number generation costs.

\begin{figure}[t]
	\centering
	\includegraphics[width=0.77\columnwidth]{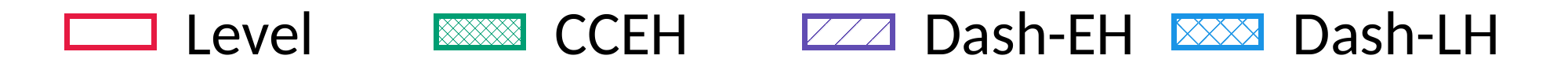}\\\vspace{-1.5mm}
	\begin{subfigure}{0.235\textwidth}
		\includegraphics[width=\columnwidth]{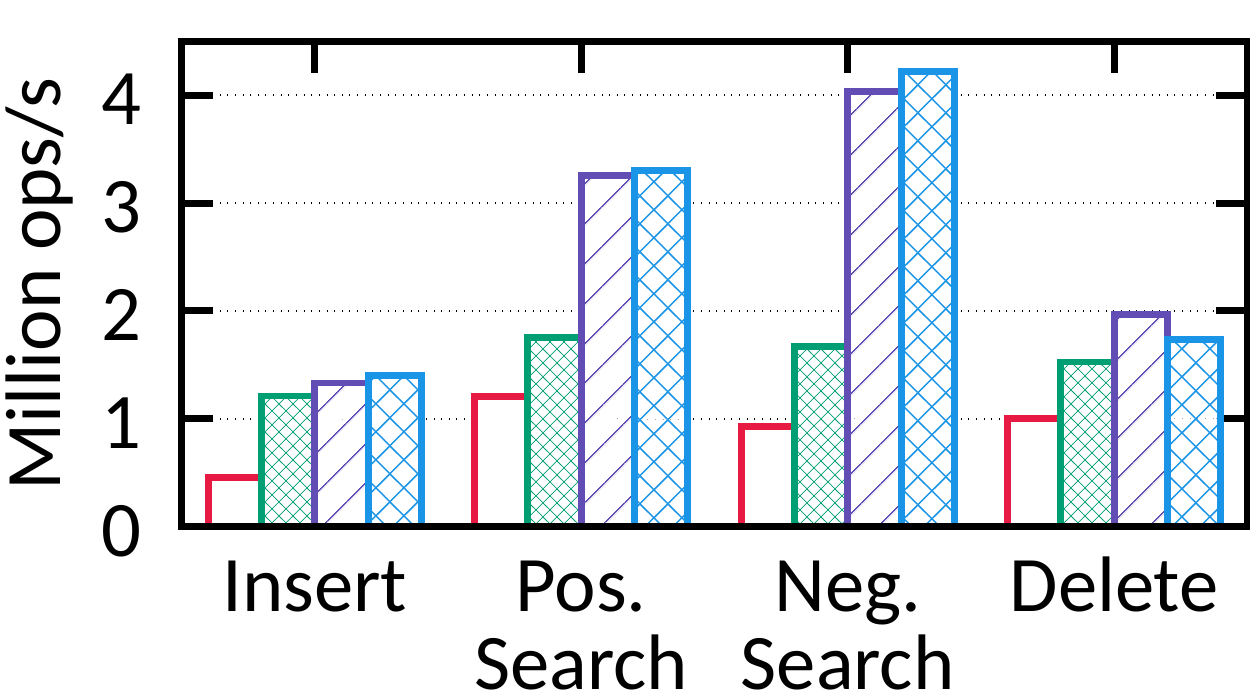}
		%\caption{Insertion for \dash-EH.}
	\end{subfigure}\hfill
	\begin{subfigure}{0.235\textwidth}
		\includegraphics[width=\columnwidth]{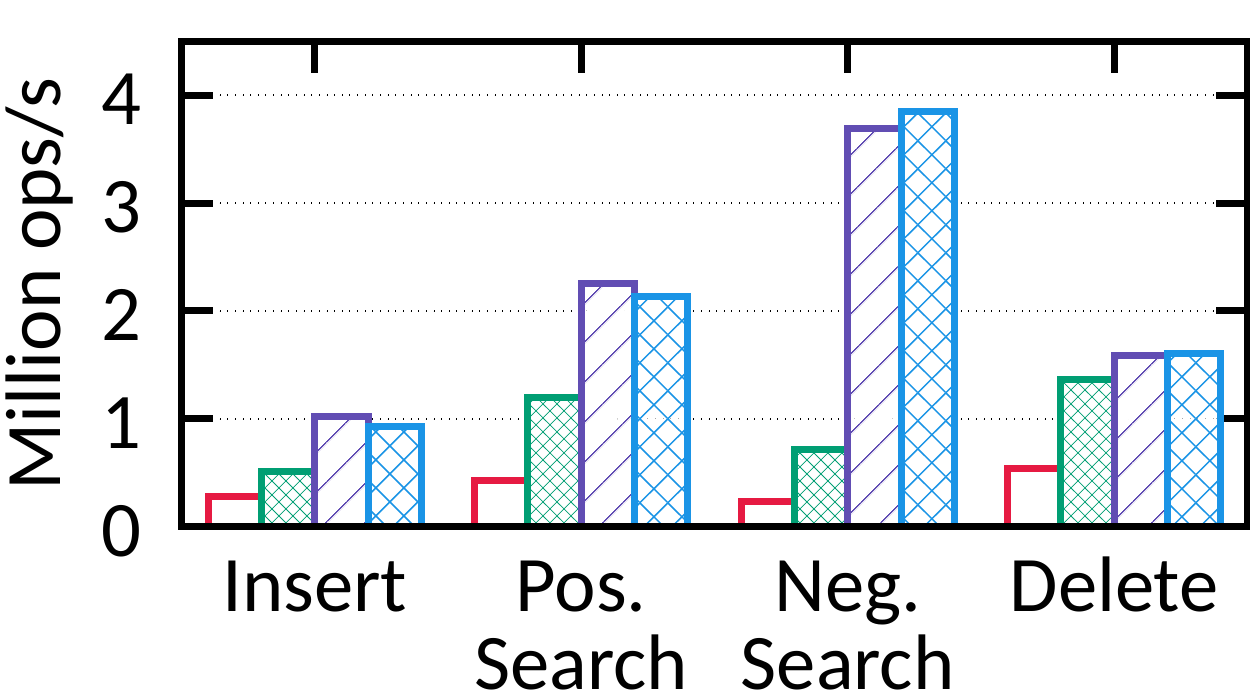}
		%\caption{Insertion for \dash-LH.}
	\end{subfigure}\hfill
	\caption{Single-thread performance under fixed-length keys (left) and variable-length keys (right).}
	\label{fig:single-throughput}
\end{figure}

\begin{figure*}[t]
\centering
\includegraphics[width=1.2\columnwidth]{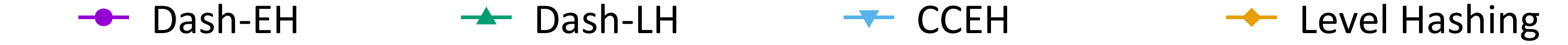}\\\vspace{-1mm}
\begin{subfigure}{0.2\textwidth}
\includegraphics[width=\columnwidth]{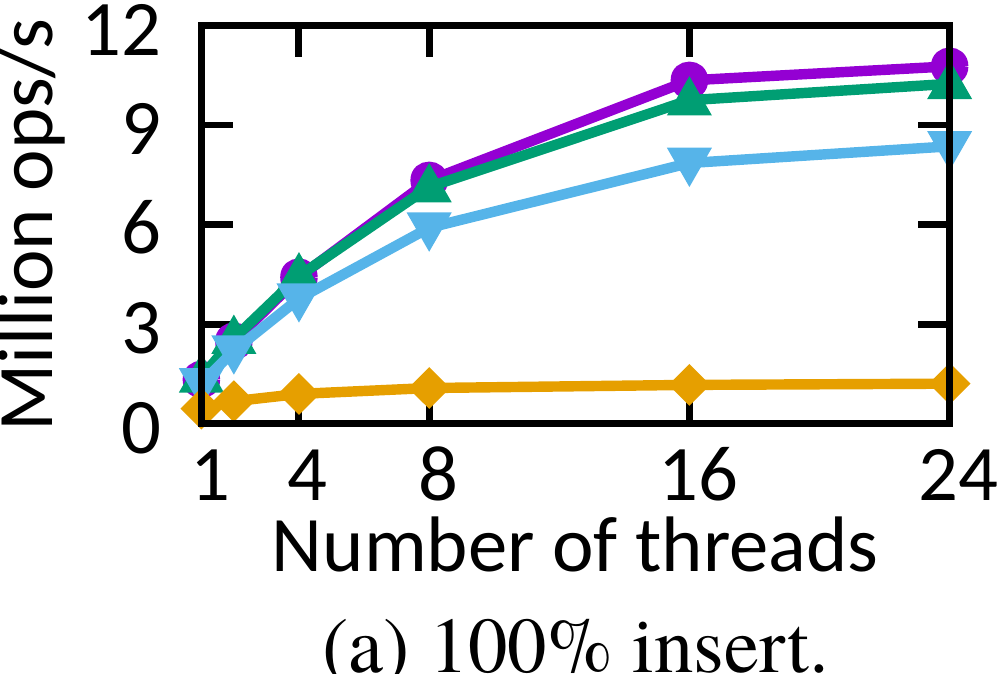}
\end{subfigure}\hfill
\begin{subfigure}{0.2\textwidth}
\includegraphics[width=\columnwidth]{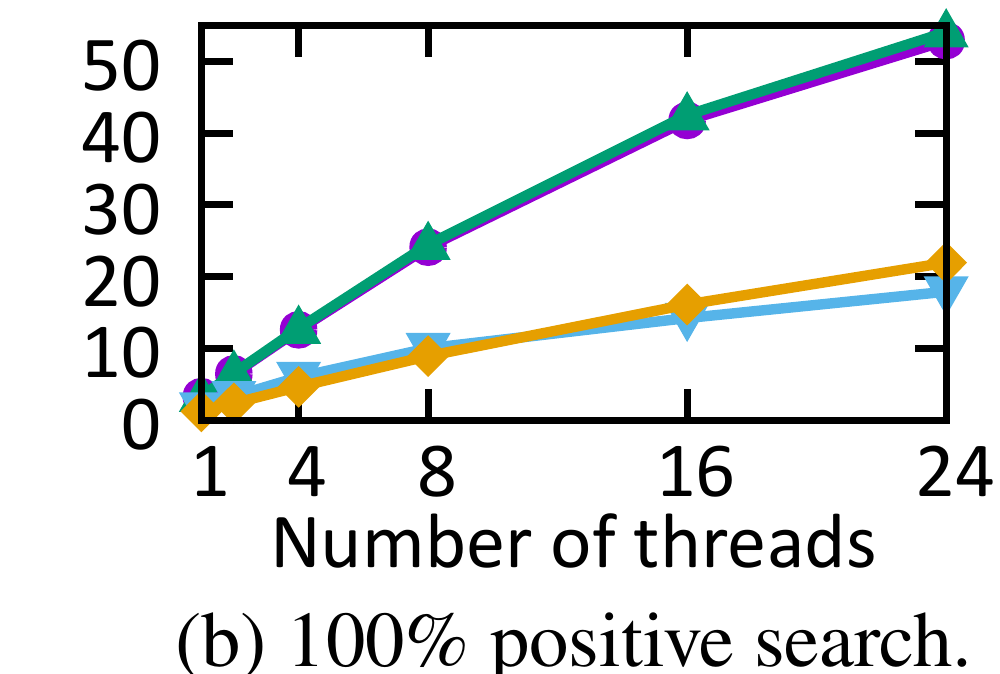}
\end{subfigure}\hfill
\begin{subfigure}{0.2\textwidth}
\includegraphics[width=\columnwidth]{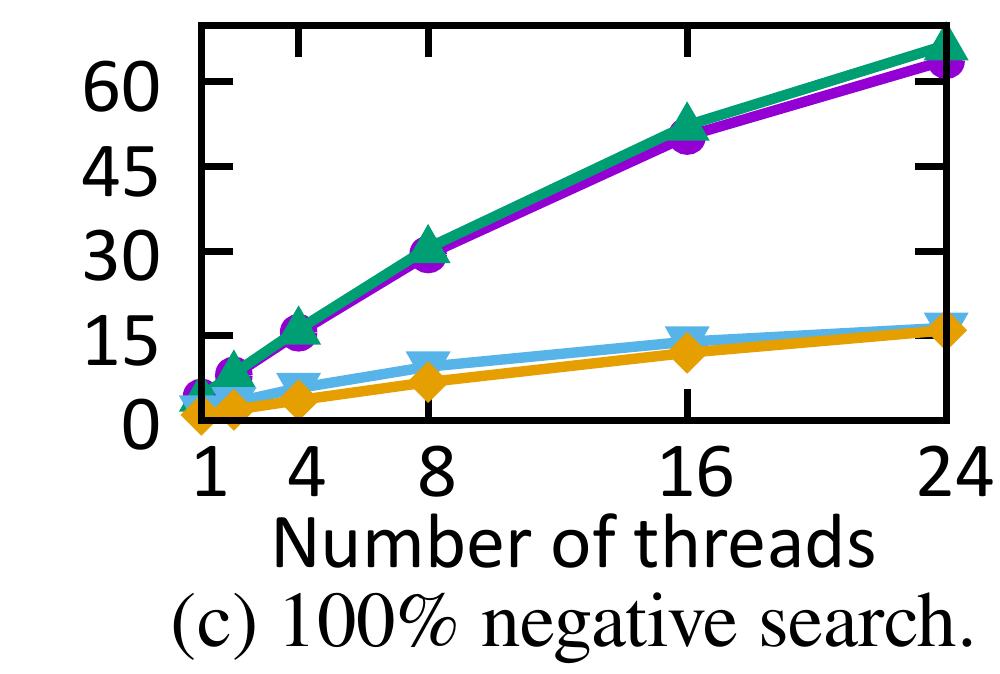}
\end{subfigure}\hfill
\begin{subfigure}{0.2\textwidth}
\includegraphics[width=\columnwidth]{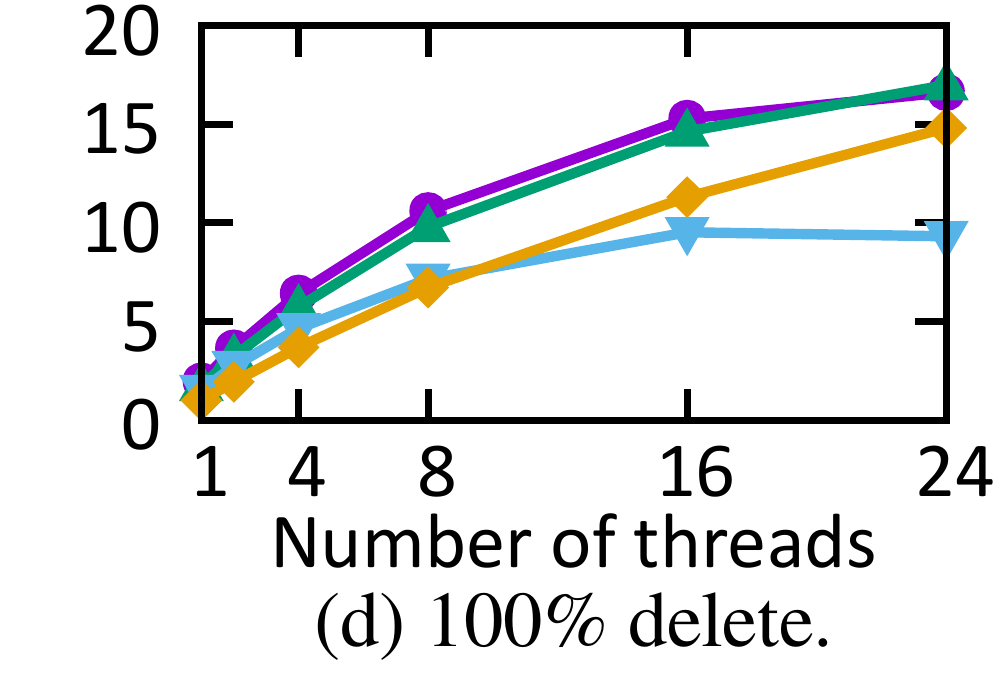}
\end{subfigure}\hfill
\begin{subfigure}{0.2\textwidth}
\includegraphics[width=\columnwidth]{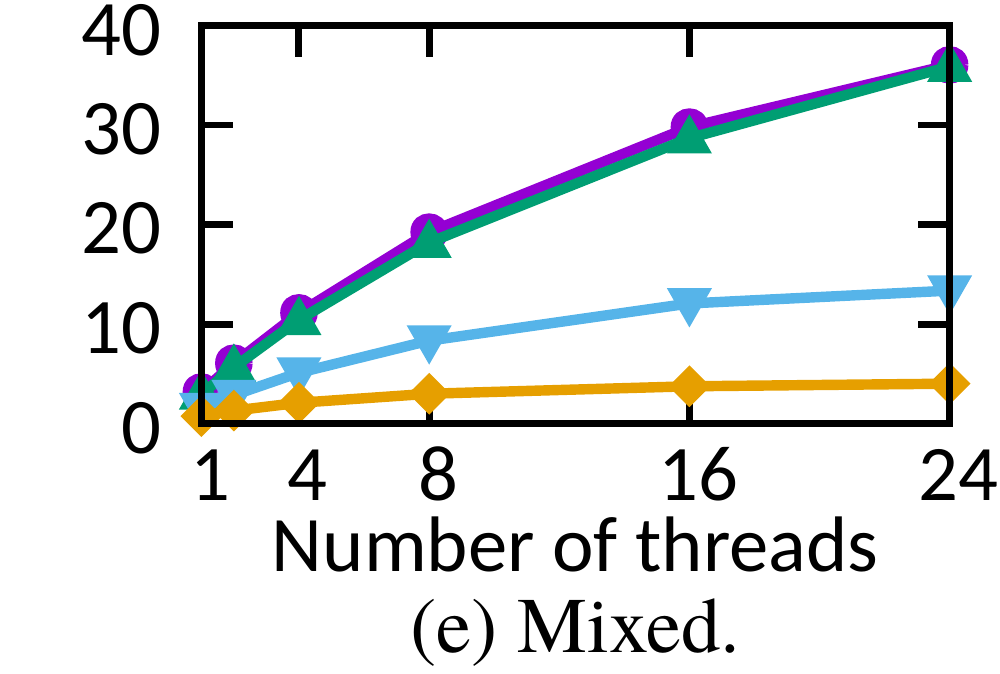}
\end{subfigure}
\caption{Throughput under different workloads with a varying number of threads and 8-byte keys and 8-byte values.}
\label{fig:fixed-length}
\end{figure*}

\subsection{Single-thread Performance}
\label{sec:rs}
We begin with single-thread performance to understand the basic behaviors of each hash table.
%In general, \dash hash tables can significantly outperform CCEH and level hashing in almost every aspect.
%More importantly, with our consistent and none ad hoc design principles,
%\dash hash tables can achieve very similar and robust performance.
%\textbf{Fixed-Length Keys.}
We first consider a read-only workload with fixed-length keys. 
Read-only results provide an upper bound performance on the hash tables since no modification is done to the data structure.
They directly reflect the underlying design's cache efficiency and concurrency control overhead.

As Figure~\ref{fig:single-throughput} shows, \dash-EH can outperform CCEH/level hashing by 1.9$\times$/2.6$\times$ for positive search.
\dash-LH and \dash-EH achieved similar performance because they use the same building blocks, with bounded PM accesses and lightweight concurrency control which reduces PM writes.
For negative search, \dash variants achieved more significant improvement, being 2.4$\times$/4.4$\times$ faster than CCEH/level hashing.
As Section~\ref{subsec:fp-meta} shows, this is attributed to fingerprints and the overflow metadata 
%(overflow bit and overflow fingerprints) 
which significantly reduce PM accesses.

For inserts, \dash and CCEH achieve similar performance ($\sim$2.5$\times$ level hashing).
Although CCEH has one fewer cacheline flush per insert than \dash, \dash's bucket load balancing strategy reduces segment splits, improving both performance and load factor.
Without the allocation bitmap, CCEH requires a reserved value (e.g., `0') to indicate an empty slot.
This design imposes additional restrictions to the application; \dash avoids it using metadata. %cannot support operations on longer (e.g. 16-byte) inlined values.
Level hashing exhibited much lower performance due to more PM reads and frequent lock/unlock operations. 
It also requires full-table rehashing that incurs many cacheline flushes.
For deletes, \dash outperforms CCEH/level hashing by 1.2$\times$/1.9$\times$ because of reduced cache misses. %line accesses.

%\textbf{Variable-Length keys.}
The benefit of \dash is more prominent for variable-length keys.
As Figure~\ref{fig:single-throughput} shows, \dash-EH/LH are 2$\times$/5$\times$ faster than CCEH/level hashing for positive search.
The differences in negative search are even more dramatic (5$\times$/15$\times$).
Again, this results show the effectiveness of fingerprinting.
Since all the hash tables store pointers for longer ($>8$-byte) keys, the accessing thread has to dereference pointers to load the keys, which is a major source of cache misses.
%This random indirection is hugely unfavored in persistent memory, 
Fingerprinting effectively reduces such indirections.
Note that all operations will benefit from this technique, because they either directly query a key (search/delete) or require uniqueness check (insert).
For the same reason, \dash-EH outperforms CCEH/level hashing by 2.0$\times$/3.7$\times$ for insert, and 1.2$\times$/2.9$\times$ for delete. 
%Figure~\ref{fig:fixed-length}a depicts the result for concurrent insert operations, 
%and Figure~\ref{fig:single-throughput} shows the throughput of single thread insert performance. 
%We observe that in the single thread case, DASH hash tables can achieve similar performance to CCEH, and they are about 2x faster than level hashing.
%With  design, the insert thread will probe the same number of cacheline as the CCEH do,
%thus these hash tables are able to achieve similar inserting performance.

%In the case of variable length inserting, however, DASH-EX is able to achieve a lot better performance than the CCEH and level hashing, 
%1.7x and 8.4x respectively as shown in Figure~\ref{fig:variable-length}a.
%This performance gap is realized by the virtue of the reduction in key dereference when performing uniqueness check.
%With the help of fingerprints, the insert thread will not need to dereference every key in the bucket, instead,
%it can check the fingerprints first, and only loads the necessary keys thus reduced a lot of cache misses.

%\begin{figure}[h]
%	\centering
%	\includegraphics[width=\columnwidth]{figures-src/pre-allocation/5-allocator.pdf}
%	\caption{Compare how the allocation policy will impact the scalability of \dash-EH(left) and \dash-LH(right).}
%	\label{fig:allocator}
%\end{figure}

\subsection{Scalability}
\label{subsec:scale}
We test both individual operations and a mixed workload that consists of 20\% of insert and 80\% of search operations. 
For the mixed workload, we preload the hash table with 60 million records to allow search operations to access actual data.
%To avoid scheduling overhead, we bound the working threads to physical cores and disabled runtime thread migration. 
%During the experiments, we disabled the hyper-threading and made sure no workload will over-subscribe the CPU cores.

%We observe that all hash tables do not scale ideally on persistent memory when performing the insert operation, and the reasons vary. 
%We observed that the most significant impact factor is memory bandwidth. 
%Figure \todo{we need to show bandwidth or comparison on DRAM} shows the same experiment running on DRAM, 
%with much larger bandwidth, the scaling factor of DASH-EH increases to  $\times$ on 24 threads. 
%For a similar reason, CCEH cannot scale on persistent memory, 
%but it also has worse scalability than \dash-EH. 
%We attribute this slow down to having more hash table splitting in CCEH. 

Figure ~\ref{fig:fixed-length} plots how each hash table scales under a varying number of threads and fixed-length keys.
%\sout{As core count increases, the throughput (million operations per second) of both CCEH and level hashing stop growing after 8 and 16 threads, respectively.}
%\textcolor{BrickRed}{
For inserts, level hashing exhibits the worst scalability mainly due to full-table rehashing, which is time-consuming on PM and blocks concurrent operations. % until it is finished.
%\dash-EH and \textcolor{BrickRed}{CCEH} scale much better, but do not do so linearly for insert workloads that exhibit many random PM writes which are inherent to hash tables.
With fingerprinting and bucket load balancing, \dash finishes uniqueness checks quickly and triggers fewer SMOs, with fewer PM accesses and interactions with the PM allocator.
Though neither \dash-EH nor \dash-LH scales linearly as inserts inherently exhibit many random PM writes, \dash is the most scalable solution, being up to 1.3$\times$/8.9$\times$ faster than CCEH/level hashing for insert operations.
%}
%We observed that the most significant impact factor is still the limited PM performance for randome writes~\cite{UCSDGuide}. 
%\dash-EH, for example, achieves 10.9$\times$ single-thread performance under 24 threads on PM. 
%However, it achieves linear scalability on DRAM for the same experiment (not shown in the figure because of limited space). 
%Section \ref{text:load-factor} will discuss more on techniques \dash hash tables used to improve load-factors and reduce memory footprint. 

For search operations, Figures~\ref{fig:fixed-length}(b--c) show near-linear scalability for \dash-EH/LH. %, which shows the effectiveness of fingerprinting and optimistic concurrency.
CCEH falls behind mainly because of its use of pessimistic locking which incurs large amount of PM writes even for read-only workloads (to acquire/release read locks).
%The recent study \cite{UCSDGuide} shows that persistent memory, specifically Intel Optane DCPMM, 
%has limited scalability on mixed read and write on the same memory region, 
%while CCEH has a read-write lock on each segment and even a search operation needs a mixed read-write on the lock. 
%The recent study \cite{UCSDGuide} shows that persistent memory, specifically Intel Optane DCPMM, 
%has limited scalability on writes (especially random writes) because of its internal contention, while CCEH has a read-write lock on each segment and even a search operation needs the writes on the lock. 
%Level hashing uses a similar design with lock striping~\cite{Mauricebook} to co-locate all the locks in a small and contiguous memory region so that they are likely to fit into the CPU cache.
Level hashing uses a similar design but lock striping~\cite{Mauricebook} makes all the locks likely to fit into the CPU cache.
Therefore, although level hashing has lower single-thread performance than CCEH, it still achieves similar performance to CCEH under multiple threads.
Delete operations in \dash-EH, \dash-LH, CCEH and level hashing on 24 threads scale and improve over their single-threaded version by 8.4$\times$, 9.8$\times$, 6.1$\times$ and 14.7$\times$, respectively.
For the mixed workload on 24 threads, \dash outperforms CCEH/level hashing by 2.7$\times$/9.0$\times$. % under fixed-length keys.

We observed similar trends (but with widening gaps between \dash-EH/LH and CCEH/level hashing) for workloads using variable-length keys (not shown for limited space). %, as shown in Figure~\ref{fig:variable-length}, 
In the following sections, we discuss how each design in \dash impacts its performance.
%we observe slightly better scalability of \dash-EH/LH and more significant performance gaps between \dash hash tables and CCEH/level hashing because of the filtering effect of the fingerprint technique. 
%\begin{figure}[t]
%	\centering
%	\includegraphics[width=1.2\columnwidth]{figures-src/pre-allocation/legend_preallocate.pdf}\vspace{-1mm}
%	\begin{subfigure}{0.23\textwidth}
%		\includegraphics[width=\columnwidth]{figures-src/pre-allocation/eh-preallocate.pdf}
		%\caption{Insertion for \dash-EH.}
%	\end{subfigure}\hfill
%	\begin{subfigure}{0.23\textwidth}
%		\includegraphics[width=\columnwidth]{figures-src/pre-allocation/lh-preallocate.pdf}
		%\caption{Insertion for \dash-LH.}
%	\end{subfigure}\hfill
  %\vspace{-3mm}
%	\caption{Impact of PM allocator; \dash-EH (left), \dash-LH (right).}
%	\label{fig:allocator}
%\end{figure}

%\begin{figure}
%	\begin{center}
%		\includegraphics[width=\columnwidth]{figures-src/read_opt/5-read_opt.pdf}
%	\end{center}
%	\caption{\label{fig:5-read_opt} Performance increase by using overflow metadata under fixed-length workload. 
%	The $two$ and $four$ prefix respectively indicates the performance when using two and four stash buckets in each segment}
%\end{figure}
%% %---------------------------
\begin{figure}[t]
	\centering
	\includegraphics[width=0.7\columnwidth]{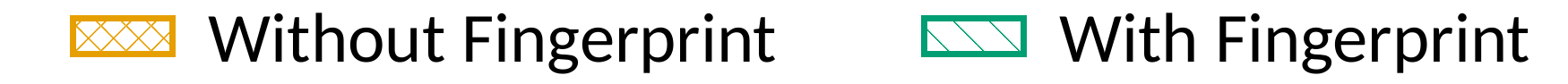}\\\vspace{-1mm}
	\begin{subfigure}{0.235\textwidth}
		\includegraphics[width=\columnwidth]{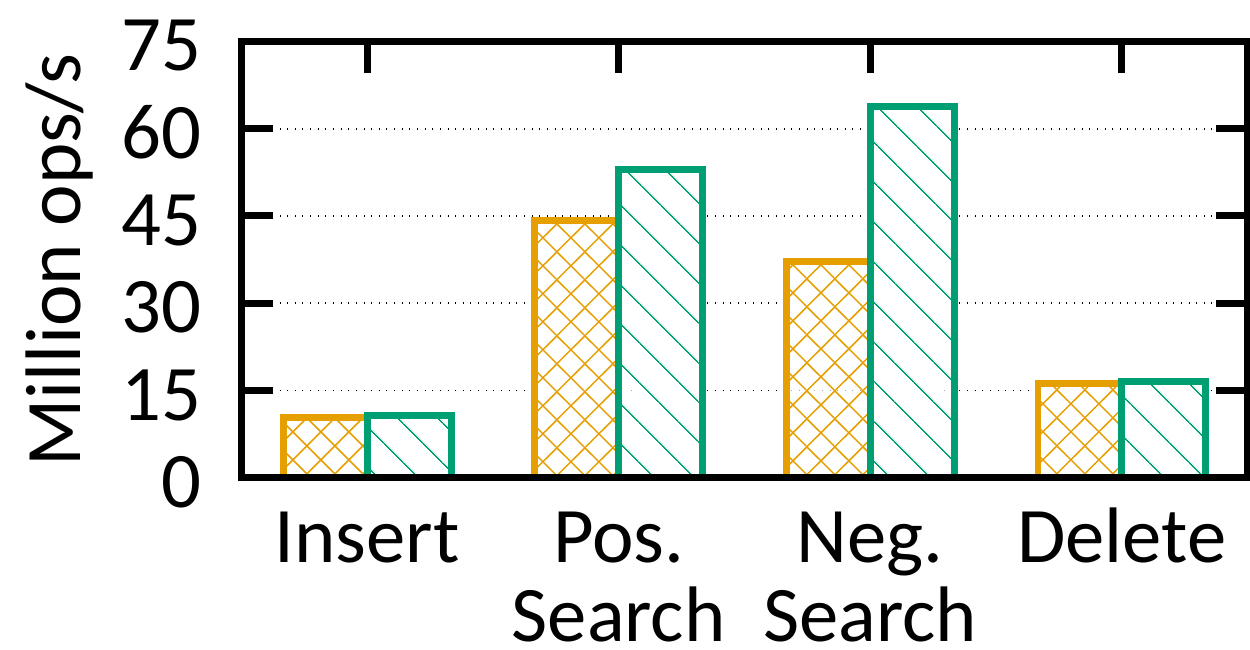}
		%\caption{Insertion for \dash-EH.}
	\end{subfigure}\hfill
	\begin{subfigure}{0.235\textwidth}
		\includegraphics[width=\columnwidth]{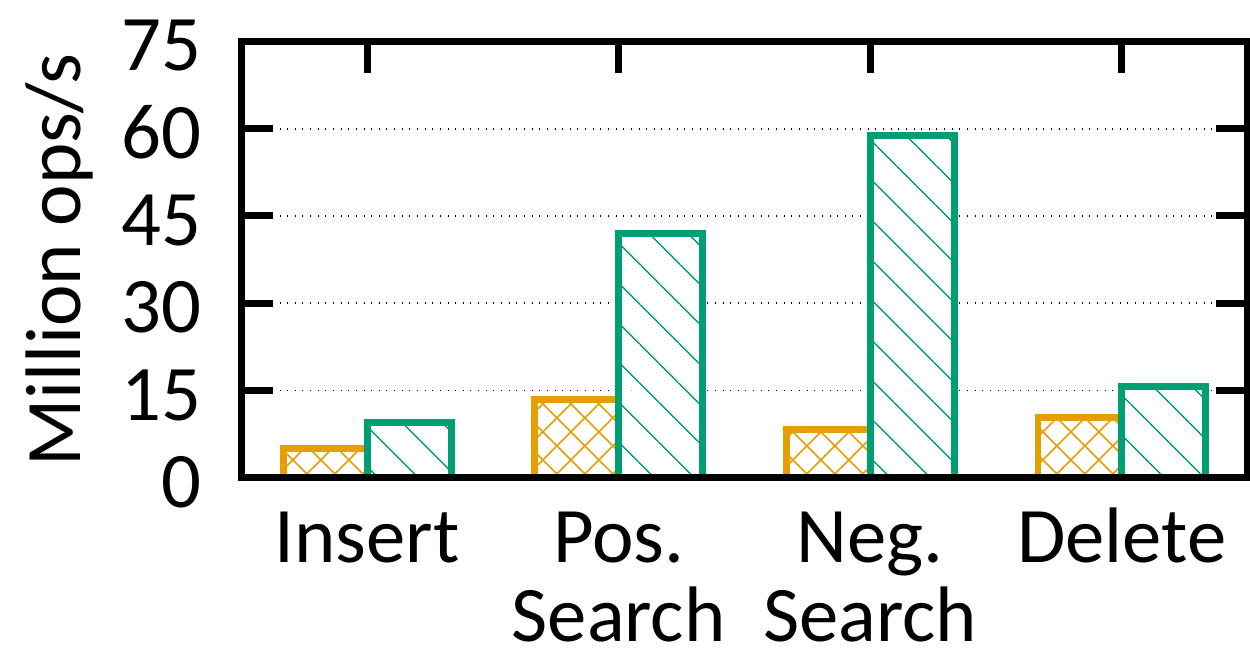}
		%\caption{Insertion for \dash-LH.}
	\end{subfigure}\hfill
	\caption{\label{fig:5-finger} Effect of fingerprinting in buckets under fixed-keys (left) and variable-length keys (right).}
\end{figure}

\begin{figure}[t]
	\centering
	\includegraphics[width=0.7\columnwidth]{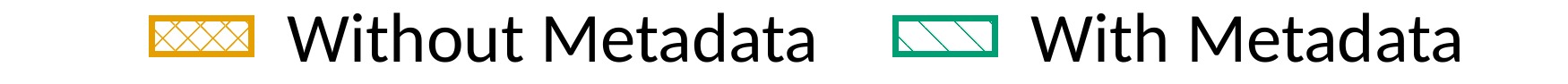}\\\vspace{-1mm}
	\begin{subfigure}{0.235\textwidth}
		\includegraphics[width=\columnwidth]{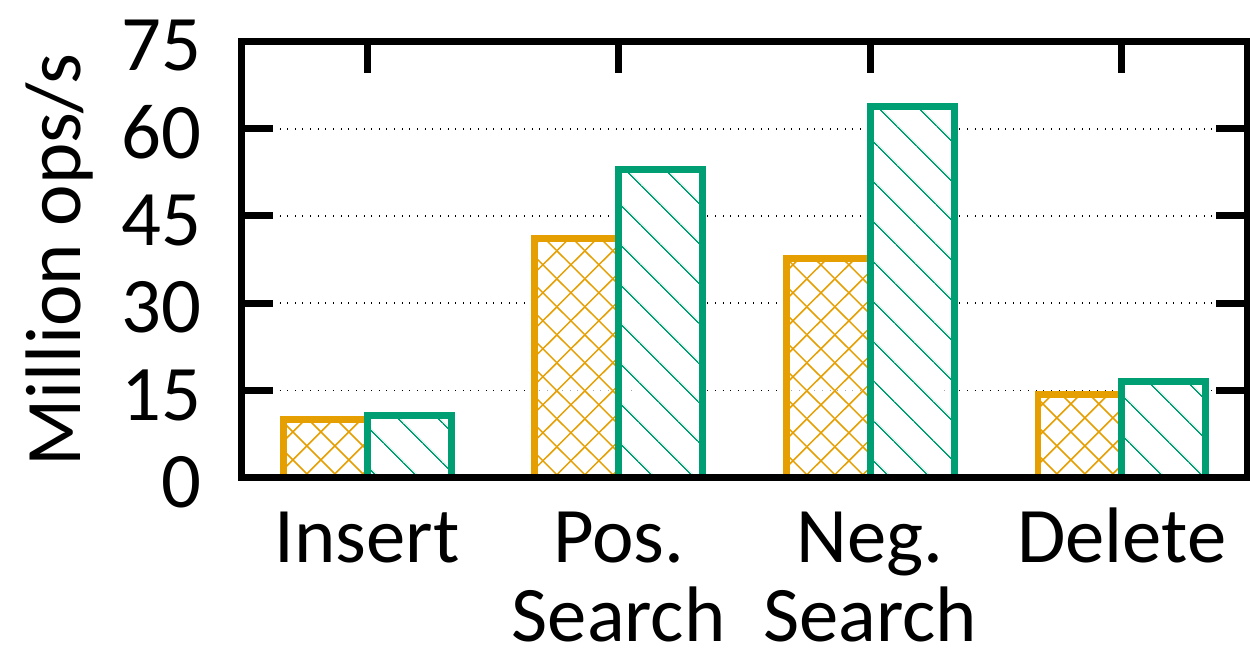}
		%\caption{Insertion for \dash-EH.}
	\end{subfigure}\hfill
	\begin{subfigure}{0.235\textwidth}
		\includegraphics[width=\columnwidth]{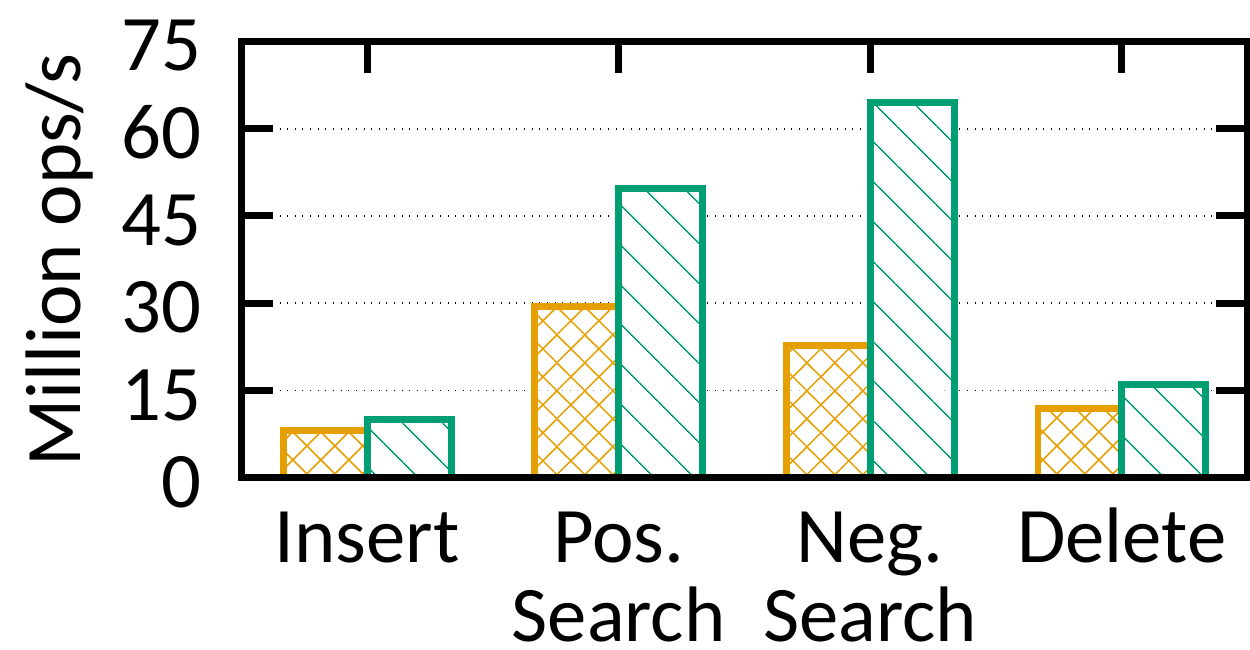}
		%\caption{Insertion for \dash-LH.}
	\end{subfigure}\hfill
	\caption{\label{fig:5-read_opt} Effect of overflow metadata (24 threads) with fixed-length keys and two (left) and four (right) stash buckets per segment.}
\end{figure}

\subsection{Fingerprinting and Overflow Metadata}
\label{subsec:fp-meta}
Fingerprinting is a major reason for \dash to perform and scale well on PM as it greatly reduces PM accesses.
We quantify its effect by comparing \dash-EH with and without fingerprinting.
%We run the -length we compare the \dash-EH with and without fingerprint for both fixed-length workload and variable-length workload with 24 threads.
Figure~\ref{fig:5-finger} shows the result under 24 threads.
With fixed-length keys, fingerprinting improves throughput by 1.04/1.19/1.72/1.02$\times$ for insert/positive search/negative search/delete.
The numbers for variable-length keys are 1.88/3.13/7.04/1.52$\times$, respectively.
%This improvement indicates that fingerprint in \dash design can help reducing the cacheline accesses and efficiently supporting the variable-length key.
As introduced in Section~\ref{subsec:space}, \dash uses overflow metadata %(overflow bit and fingerprints for records stored in stash buckets) 
to allow early-stop of search operations to save PM accesses.
%The metadata shows whether the stash buckets has the search record and if so, which stash bucket owns the records.
Figure~\ref{fig:5-read_opt} shows its effectiveness under 24 threads with varying numbers of stash buckets.
\dash-EH with two stash buckets outperforms the baseline (no metadata) by 1.07/1.29/1.70/1.16$\times$ for insert/positive search/negative search/delete.
%We observe that this method significantly enhanced search performance because of the reduce of memory accesses.
%We also noticed that, with more stash buckets configured, the search performance dropped about 25\% without metadata, 
With more stash buckets added, search performance drops by about 25\% without the overflow metadata.
With overflow metadata, however, the performance remains stable, as negative search operations can early-stop after checking the overflow metadata without actually probing the stash buckets.
%This property indicates the robustness of \dash design; 
%it holds the leadership under different configurations.

\begin{figure}[t]
	\centering
	\includegraphics[width=0.93\columnwidth]{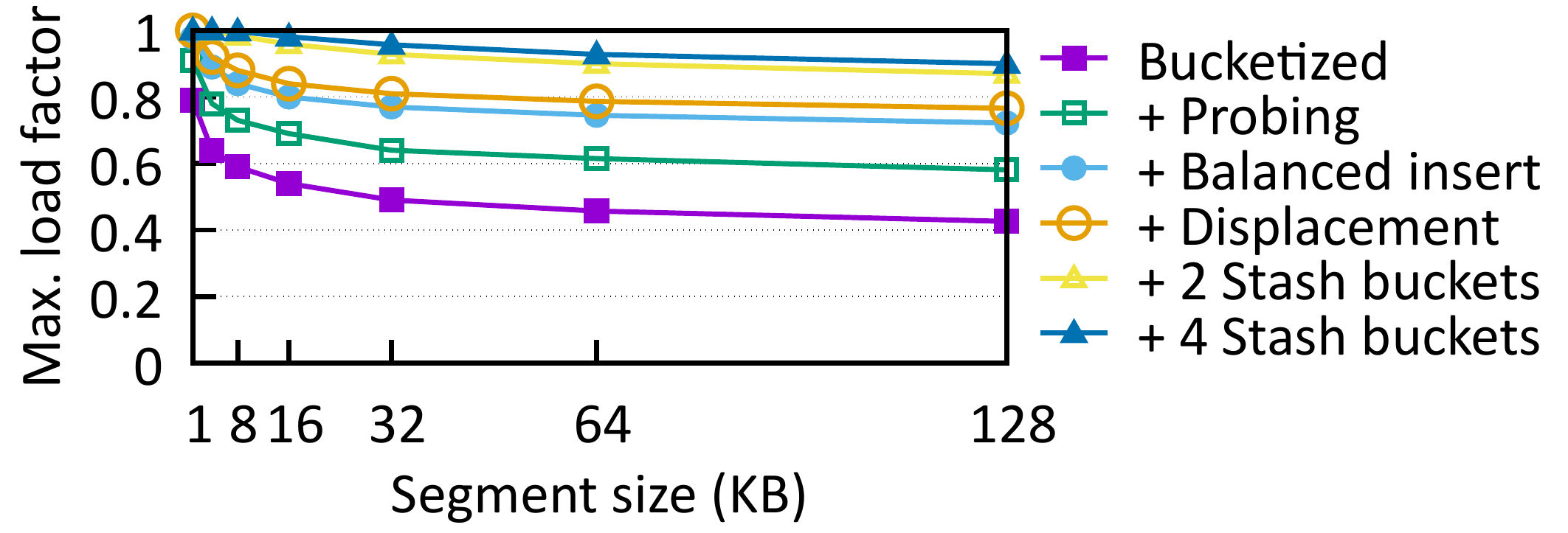}
  %\vspace{-6mm}
	\caption{Maximum load factor after adding different techniques.}
	\label{fig:load-factor-analysis}
\end{figure}

\subsection{Load Factor}
\label{subsec:load-factor}
Now we study how linear probing, balanced insert, displacement and stashing improve load factor.
We first measure the maximum load factor of one segment with different sizes.
Using larger segments could decrease load factor, but may improve performance by reducing directory size.
Real systems must balance this tradeoff. %, though it is desirable to have both high load factor and large segments.

Figure~\ref{fig:load-factor-analysis} shows the result. % under different techniques across varying segment size settings.
In the figure ``Bucketized'' represents the simplest segmentation design without the other techniques.
The segment can be up to 80\% full on 1KB segments but gradually degrades to about 40\% full at most as the segment size increases to 128KB.
With just one bucket probing added (``+Probing''), the load factor under 128KB segment increases by $\sim$20\%. 
Balanced insert and displacement can improve load factor by another $\sim$20\%.
With stashing, we maintain close to 100\% load factor for 1--16KB segments.
\dash combines all these techniques to achieve more than twice the load factor than vanilla segmentation with large segments. 
%For all later performance evaluations, we set the segment size to 16KB, 
%and each of the segments has two stash buckets. 
%These parameters are tunable as per real-world scenarios.

\begin{figure}[t]
	\centering
	\includegraphics[width=0.98\columnwidth]{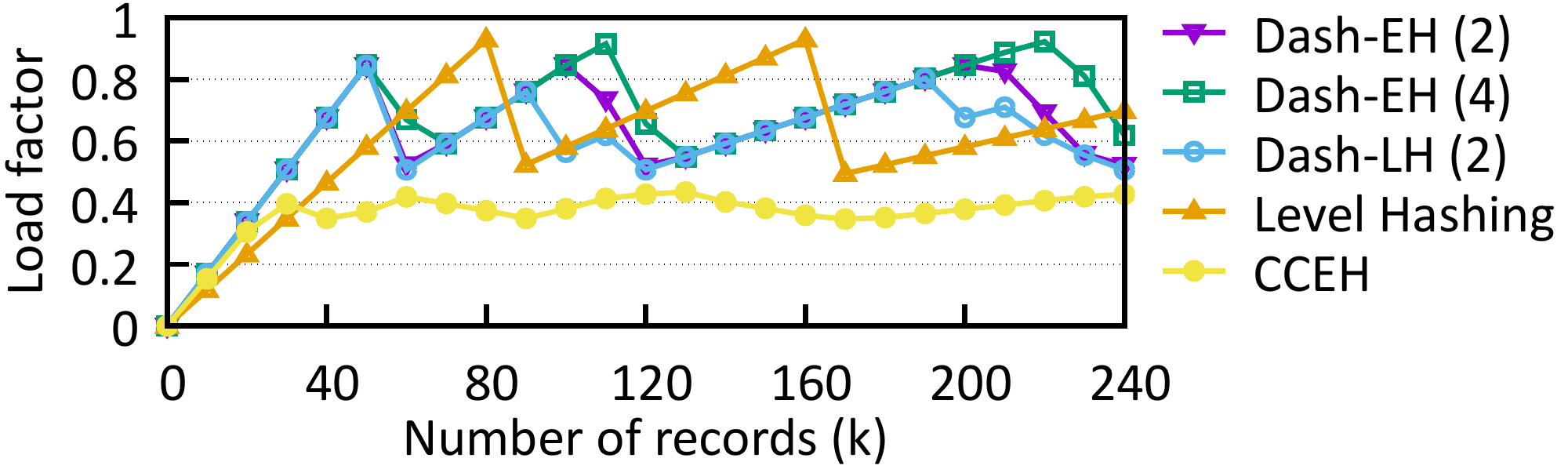}
  %\vspace{-3mm}
	\caption{Load factor of different hashing schemes with respect to number of items inserted to the hash table.}
	\label{fig:load-factor-compare}
\end{figure}

To compare different designs more realistically, we observe how load factor changes after a sequence of inserts.
We start with an empty hash table (load factor of 0) and measure the load factor after different numbers of records have been added to the hash tables.
As shown in Figure~\ref{fig:load-factor-compare}, the load factor (y-axis) of CCEH fluctuates between 35\% and 43\%, because %. % over the x-axis. 
CCEH only conducts four cacheline probings before triggering a split.
%We also found a bug in the open-source code of CCEH, where they report higher load factors (i.e., from 50\% to 60\%) than the actual values. 
%The CCEH authors acknowledged this bug in their GitHub issues\footnote{https://github.com/DICL/CCEH/issues/2}. 
As we noted in Section~\ref{subsec:space}, long probing lengths increase load factor at the cost of performance, yet short probing lengths lead to pre-mature splits.
Compared to CCEH, \dash and level hashing can achieve high load factor because of their effective load factor improvement techniques. 
The ``dipping'' indicates segment splits/table rehashing is happening. 
We also observe %the impact of the number of stash buckets on load factor.
that with two stash buckets, denoted as \dash-EH/LH~(2), we achieve up to 80\% load factor, while the number for using four stash buckets in \dash-EH (4) is 90\%, matching that of level hashing.
%Note that both level hashing and \dash-EH/LH use additional space to store metadata. 
%However, the metadata is small enough since both the bucket of level hashing and \dash hash tables only uses 12.5\% space to store the metadata.
%Moreover, \dash tables make full use of the metadata space while level hashing only stores the bitmap and wastes remaining space due to the padding bits for cacheline alignment. 
%---------------------------

\begin{figure}[t]
	\begin{center}
		\includegraphics[width=0.93\columnwidth]{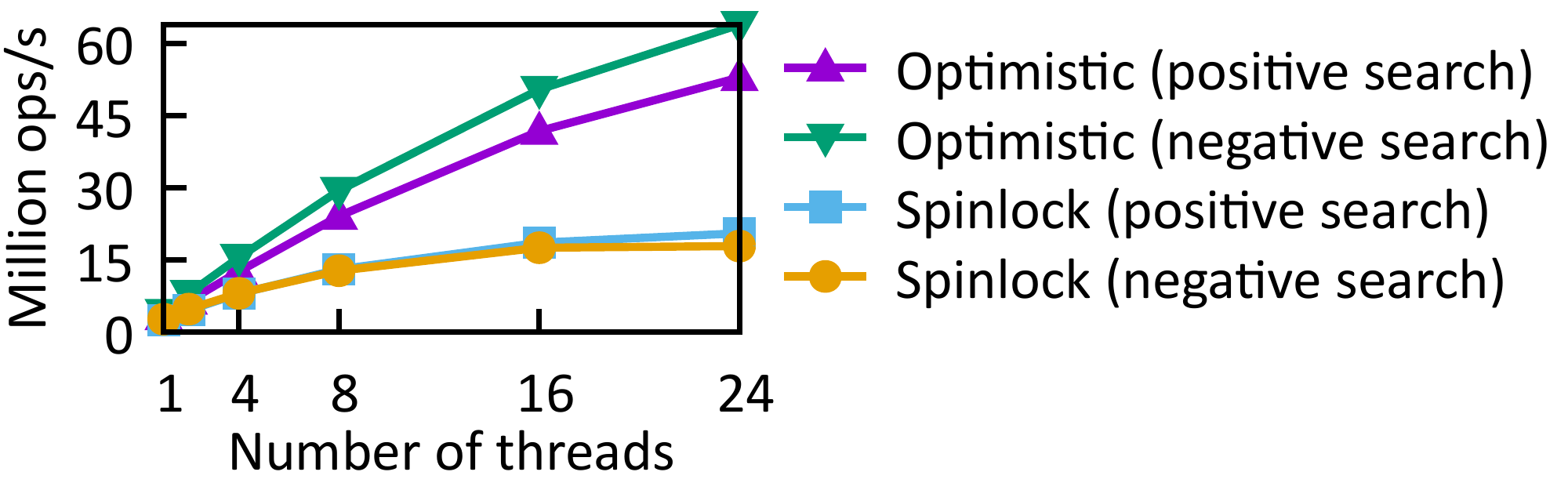}
	\end{center}
  %\vspace{-4mm}
	\caption{\label{fig:lock} Scalability under different concurrency control strategies (reader-writer spinlocks vs. optimistic locking in \dash-EH).}
\end{figure}

\subsection{Impact of Concurrency Control}
%PM's lower bandwidth and higher latency make it important for concurrent PM data structures to choose a suitable concurrency control protocol that does not incur excessive PM accesses.
As Section~\ref{subsec:cc} describes, PM data structures favor lightweight approach such as optimistic locking in \dash, over traditional pessimistic locking.
%Figure~\ref{fig:lock} shows that the \dash with optimistic locks can outperform \dash with read-write spinlocks by at least twice.
Figure~\ref{fig:lock} experimentally verifies this point by comparing pessimistic locking (reader-writer spinlocks) and optimistic locking in \dash-EH, under (negative and positive) search workloads.
%Under both negative and positive search workloads, 
The spinlock based version does not scale well because of extra PM writes needed for manipulating read locks.
We repeated the same experiments on DRAM and found that both of them can scale well.
This captures yet another important finding that was omitted or not easy to discover in previous emulation based studies.
%their performance might significantly diverge on real persistent memory, 
%and previous simulated studies failed to capture these distinctions.

%The most notable observation from the results shown in Figure~\ref{fig:load-factor-compare} is that the load factor changes periodically,

\subsection{Recovery}
It is desirable for persistent hash tables to recover instantly after a crash or clean shutdown to reduce service downtime.
We test recovery time by first loading a certain number of records and then killing the process and measuring the time needed for the system to be able to handle incoming requests.
%Note that \dash hash tables and CCEH require a dedicated recovery process to ensure data integrity. 
%Particularly, \dash hash tables require an initial recovery and a lazy metadata rebuilding to reach full performance; CCEH requires a directory scanning process.
%We first test the recovery time for all hash tables of different workload sizes before they can serve a new normal operation.
Table~\ref{tab:recovery-time} shows the time needed for each hash table to get ready for handling incoming requests under different data sizes.
The recovery time for \dash-EH/LH and level hashing are at sub-second level and does not scale as data size increases, effectively achieving instant recovery.
%Neither \dash nor level hashing requires lengthy recovery work to be done upon recovery.
For \dash-EH/LH the only needed work is to open the PM pool that is backing the hash table, and then read and possibly set the values of two variables. 
For 1280M data, level hashing requires an allocation size greater than the maximum allowed by PMDK allocator (15.998GB). % so we had to omit the experiment for level hashing under large data sets (e.g., 1280 million and beyond).
However, it also only needs a fixed amount of work to open the PM pool during the recovery, so we expect its recovery time would remain the same (53ms) under larger data sizes.
%We also observed that the recovery time of \dash-LH, given its extra directory scanning step, is nearly the same as the level hashing; this is because \dash-LH used hyrbid expansion in segment splitting thus maintained logarithmic directory size to the workloads. 
The recovery time for CCEH is linearly proportional to the size of the indexed data because it needs to scan the entire directory upon recovery.
As data size increases, so is the directory size, requiring more time on recovery.

\begin{figure}[t]
	\centering
  \includegraphics[width=0.49\columnwidth]{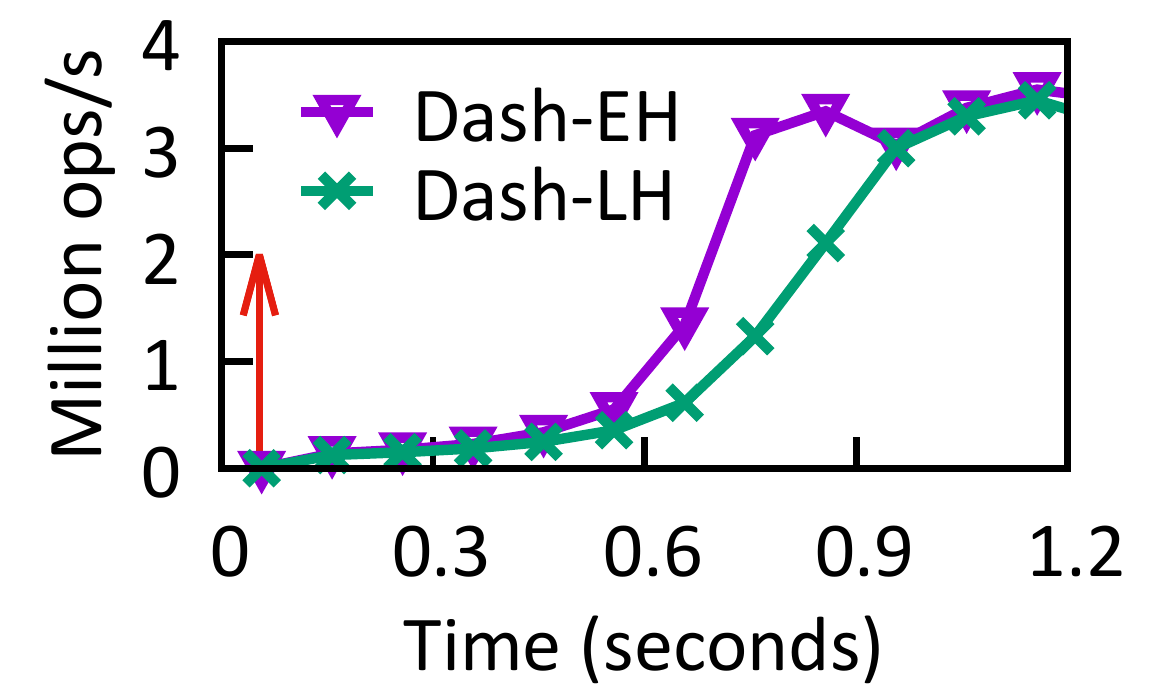}
  \includegraphics[width=0.49\columnwidth]{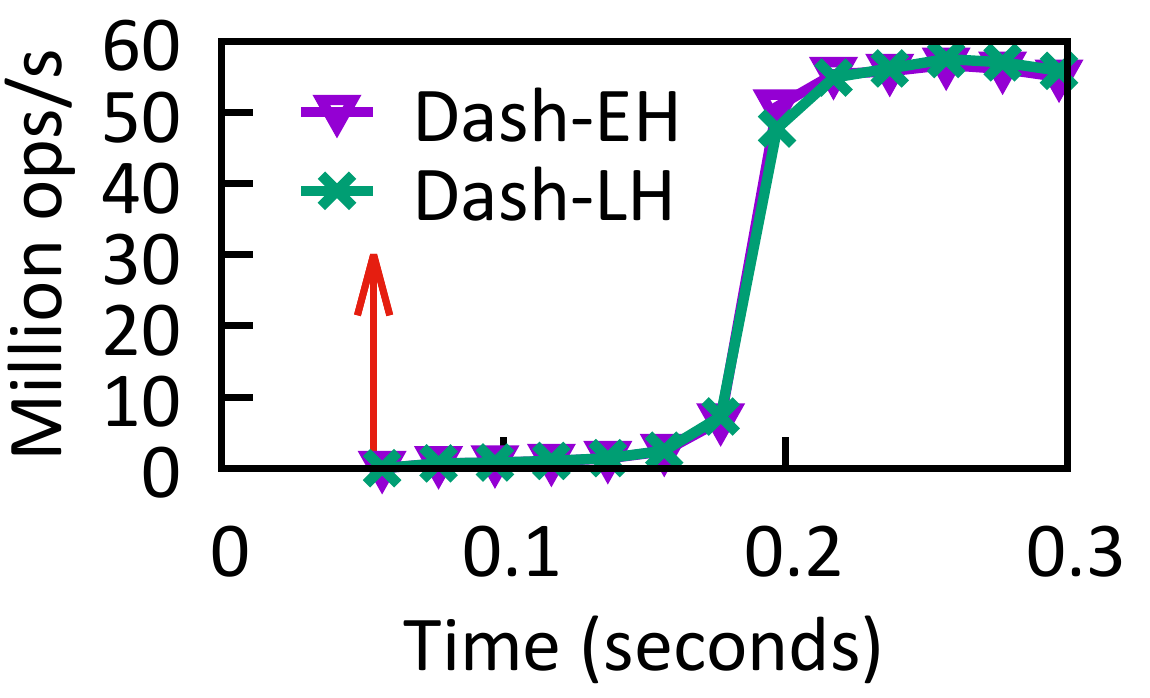}
		\caption{Throughput under different time points upon restart with one thread (left)
    and 24 threads (right). 
    %Recovery work is amortized over requests handling; after most
    %segments are recovered as part of normal accesses, throughput gradually
    %recovers back to normal.
    }
	\label{fig:lazy-recovery}
\end{figure}

%After the initial recovery, \dash starts to serve requests and simultaneously rebuilds metadata when needed. 
\dash's lazy recovery may impact runtime performance. %, for instance if an on-going insert hits a segment that needs recovery.
We measure this effect by recording the throughput over time of \dash-EH/LH once it is instantly recovered.
The hash table is pre-loaded with 40 million records.
We then kill the process while running a pure insert workload and restart the hash table and start to issue positive search operations to observe how throughput changes over time.
The result is shown in Figure~\ref{fig:lazy-recovery}; the red arrow indicates the time point when \dash is back online to be able to serve new requests.
The throughput is relatively low at the beginning: 0.1--0.3 Mops/s under one thread in Figure~\ref{fig:lazy-recovery}(left), and 0.6 Mops/s under 24 threads in Figure~\ref{fig:lazy-recovery}(right).
%(2) the threads have a high probability of hitting non-recovered segments and thus need to scan the segment and rebuild the metadata.
%As shown in Figure~\ref{fig:lazy-recovery}(right), 
Using more threads can help throughput to return to normal earlier, as multiple threads could hit different segments and work on the rebuilding of metadata or concluding SMOs in parallel.
Throughput returns to normal in 0.2 seconds under 24 threads, while the number under one thread is 0.9 s.
\begin{table}[t]
	\centering
	\caption{Recovery time (ms) vs. data size. CCEH's recovery time scales with data size. For level hashing and \dash it remains constant because both need a fixed amount of work upon restart.}
  %For level hashing we show the expected value for 1280M, as it requires an allocation size that exceeds the maximum supported by PMDK allocator.}}
  %\vspace{-3mm}
	\begin{tabular}{c|c|c|c|c|c|c}
  \hline
		\multirow{2}{*}{Hash Table} &
		\multicolumn{6}{c}{\bf Number of indexed records (million)} \\ \cline{2-7}
		& 40 & 80 & 160 & 320 & 640 & 1280  \\ \hline
		\bf \dash-EH     	& 57   & 57   & 57   & 57  & 57 & 57    \\ 
		\bf \dash-LH     	& 57   & 57   & 57   & 57  & 57 & 57    \\ 
		\bf CCEH        	& 113    & 165    & 262   & 463  & 870	& 1673     \\ 
		\bf Level hashing   & 53    & 53   & 53   & 53  & 53  & (53)  \\
    \hline
	\end{tabular}
	\label{tab:recovery-time}
\end{table}

\subsection{Impact of PM Software Infrastructure}
\label{subsec:pm-infra}
\begin{figure}[t]
\centering
\includegraphics[width=\columnwidth]{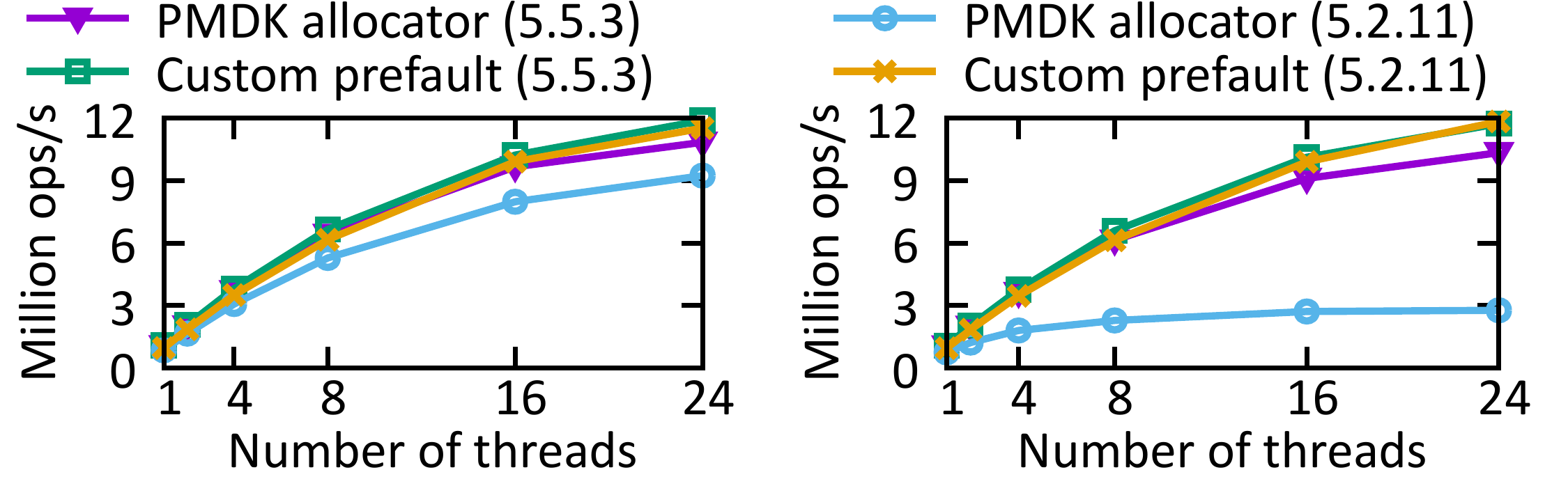}
%\vspace{-3mm}
\caption{Impact of PM allocator and OS support on \dash-EH (left) and \dash-LH (right).}
\label{fig:allocator}
\end{figure}

It has been shown that PM programming infrastructure can be a major overhead due to reasons such as page faults and cacheline flushes~\cite{PiBench,UCSDMeasurement}. % on the critical path.
%In the process of implementing \dash, we also found lower level, kernel support can play a role in impacting performance. 
We quantify its impact by running the same insert benchmark in Section~\ref{subsec:scale} under two allocators (PMDK vs. a customized allocator) and two Linux kernel versions (5.2.11 vs. 5.5.3).
Our customized allocator pre-allocates and pre-faults PM to remove page faults at runtime.
Though not practical, it allows us to quantify the impact of PM allocator; it is not used in other experiments.
As Figure~\ref{fig:allocator}(left) shows, \dash-EH is not very sensitive to allocator performance under different kernel versions as its allocation size (single segment, 16KB) is fixed and not huge.
\dash-LH in Figure~\ref{fig:allocator}(right), however, exhibited very low performance using PMDK allocator on kernel 5.2.11 ($\sim25\%$ the number under 5.5.3).
%Note that PMDK allocator performed reasonably well on kernel 5.5.3 for both \dash variants, although \dash-LH's hybrid expansion puts more pressure on the allocator (the allocation size is increasingly larger).
%(each allocation may be $2\times$ the previous allocation size).
We found the reason was a bug\footnote{Caused by a patch discussed at \href{https://lkml.org/lkml/2019/7/3/95}{\texttt{lkml.org/lkml/2019/7/3/95}}, fixed by patch at \href{https://lkml.org/lkml/2019/10/19/135}{\texttt{lkml.org/lkml/2019/10/19/135}} (5.3.8 and newer). More details are available in our code repository.} in kernel 5.2.11 that can cause large PM allocations to fall back to use 4KB pages, instead of 2MB huge pages (PMDK default).
This led to many more page faults and OS scheduler activities, which impact \dash-LH the most, as linear hashing inherently requires multiple threads compete for PM allocation during split operations and slow PM allocation could block concurrent requests. %}
The increased number of page faults also impacted recovery performance.
For instance, under 160 million records, it took CCEH $\sim$10$\times$ longer on kernel 5.2.11 than the number in Table~\ref{tab:recovery-time} to recover.%}

%As shown in figure~\ref{fig:allocator}, with no existence of bug (i.e., on 5.5.3), pre-allocation only brings minor improvement since PM write bandwidth still dominates the performance. 
%However, on kernel 5.2.11, the insertion performance of \dash drops obviously compared with pre-allocation results. 

%uIn addition to the characteristics of PM devices, we note that PM programming infrastructure also plays an important role with performance.
%Due to the large capacity of PM, using small pages (e.g., 4KB) on a system could incur non-trivial overhead due to more kernel memory usage and traversal on the large page table. 
%To relieve this overhead, PMDK employs 2MB huge page, which could also reduce the number of page faults.
%However, we identify that a bug\footnote{Detail referring to \url{https://github.com/torvalds/linux/commit/a71d83cde91e0be74db3b1c3682ee2cab4a2f413}} in Linux kernel from version 5.1.20 to 5.3.7 could destroy this optimization and significantly impact the performance of PM data structures. 
%This bug makes the huge page mapping always fail and fall back to the normal page size during a page fault. 
%Under this bug, the PM allocation could be extremely slow with more page faults incurred. 

%\dash also took longer time to restore to the normal throughput.}
%Taking huge page bug as an example, our results highlight the importance of better PM programming infrastructure, which is promising but orthogonal to our work, and can benefit PM data structures in general.

%\textcolor{BrickRed}{
These results highlight the complexity of PM programming and call for careful design and testing involving both userspace (e.g., allocator) and OS support.
We believe it is necessary as the PM programming stack is evolving rapidly while practitioners and researchers have started to rely on them to build PM data structures.%}
%In addition to the needed lazy recovery work while handling search requests, we note that the efficiency of the underlying PM library also play a role in it.
%Upon first accesses to PM pages, we observed lots of page faults and cache misses.
%------------------ TODOs --------------------%
% tz: I would do it in the experiment section
%\todo{do we need clarify the size of the segment?}

%\todo{tz: this should go to evaluation}
%\textbf{Parameter setting.} The number of overflow fingerprints reserved in each bucket is tunable according to the proportional relationship between the number of normal buckets and stash buckets. 
%In our evaluation, if the segment has 64 normal buckets(16 KB in total assuming each bucket is 256 bytes), 2 stash buckets is enough to ensure that the segment achieves more than 95\% load factor. 
%Therefore, the average number of key-value records that are overflowed from one bucket is small. 
%Only few slots of overflow fingerprints(e.g., 4) need to be reserved in each bucket so that the space overhead is small. 
%The number of bits reserved for stash bucket index information is related the number of overflow fingerprints in one bucket and the number of stash buckets in one segment.
%As shown in Figure~\ref{fig:bucket}, the bucket reserves 8 bits between bits 216 and bits 224 in the metadata area, then each overflow fingerprint has 2 bits to record the stash index information(assume the segment has 4 stash buckets).

\section{Related Work}
\label{sec:related-work}
%\textcolor{BrickRed}{
\dash builds upon many techniques from prior in-memory and PM-based hash tables, tree structures and PM programming tools.%}

%\textcolor{BrickRed}{
\textbf{In-Memory Hash Indexes.}
Section~\ref{sec:bg-dyn-hash} has covered extendible hashing~\cite{ExtHashing} and linear hashing~\cite{LarsonLinearHashing,LitwinLinearHashing}, so we do not repeat here.
%Chained hashing~\cite{artprogramming} is a basic design to resolve hash collisions by storing conflicting records in per-bucket linked lists. 
%Another chaining based design is split-ordered list~\cite{SplitOrderedLists}, which organize records in a specific order to enables lock-free resizing and high concurrency.
%Both approaches suffer from high cache miss rates due to pointer chasing and memory allocation overheads. 
%Unlike these chaining-based designs, \dash avoids pointer chasing by limiting the number of bucket probings.
Cuckoo hashing~\cite{CuckooHashing} achieves high memory efficiency through displacement: %constant lookup cost and 
a record can be inserted into one of the two buckets computed using two independent hash functions; 
if both buckets are full, a randomly-chosen record is evicted to its alternative bucket to make space for the new record.
The evicted record is inserted in the same way. %This process is repeated for the evicted record until an empty slot is found. 
%Cuckoo hashing is efficient for lookups, but lacks efficient concurrency support.
%Cuckoo hashing is widely used because of the constant lookup cost and high memory efficiency but lacks the efficient support for concurrency.
MemC3~\cite{MemC3} proposes a single-writer, multi-reader optimistic concurrent cuckoo hashing scheme that uses version counters with a global lock. 
%Since MemC3 conducts complex cuckoo displacements, 
%but only allows one writer at a time to avoid writer-writer deadlocks. 
MemC3~\cite{MemC3} also uses a tagging technique which is similar to fingerprinting to reduce the overhead of accessing pointers to variable-length keys. 
FASTER~\cite{FASTER} further optimizes it by storing the tag in the unused high order 16 bits in each pointer.
\texttt{libcuckoo}~\cite{libcuckoo} extends MemC3 to support multi-writer. % with shorter cuckoo path length. 
Cuckoo hashing approaches may incur many memory writes due to consecutive cuckoo displacements. % and also do not ensure durability on PM. 
\dash limits the number of probings and uses optimistic locking to reduce PM
writes.%} % rather than to increase concurrency by considering the workload characteristic. 

%Compared to these techniques, \dash studies the influence of different concurrency control methods on the design of persistent hash tables on real PM.
%\textcolor{BrickRed}{
\textbf{Static Hashing on PM.} 
%Resizing a static hash table requires full-table rehashing which can block concurrent operations and degrade performance.
%Static hashing inherently requires full-table rehashing which can block concurrent operations and degrade performance.
Most work aims at reducing PM writes, improving load factor and reducing the cost of full-table rehashing.
%This is often at the price of more expensive search operations and lower space utilization.
Some proposals use multi-level designs that consist of a main table and additional levels of stashes to store records that cannot be inserted into the main table.
PFHT~\cite{PFHT} is a two-level scheme that allows only one displacement to reduce writes.
It uses linked lists to resolve collisions in the stash, which may incur many cache misses during probing. 
Path hashing~\cite{PathHashing} further organizes the stash as an inverted complete binary tree to lower search cost. 
%This allows PFHT to reduce PM writes, improve load factor and avoid frequent full-table rehashing.
%But the linked lists used to resolve hash collisions in the stash may incur many cache misses during probing. 
%It only allows one cuckoo displacement during insert, to reduce PM writes in cuckoo hashing. If there is no possible displacement path, PFHT stores the new key-value item to a stash area to improve the load factor.Stashes in PFHT are small hash tables that use linked lists per bucket to resolve collisions. Linked lists require expensive linear search that is cache-unfriendly.
%But the search cost still increases with the number of buckets in logarithmic scale. 
Level hashing~\cite{LevelHashing} is a two-level scheme that bounds the search cost to at most four buckets.
%It consists of two levels of hash tables where the bottom level is a stash for the top level. %, and both of them are cuckoo hash table that only allows one displacement. 
Upon resizing, the bottom-level is rehashed to be 4$\times$ size of the top-level table, and the previous top level becomes the new bottom level. %on the role of the bottom-level stash in the new hash table. 
Compared to cuckoo hashing, the number of buckets needed to probe during a lookup is doubled.
\dash also uses stashes to improve load factor, but most search operations only need to access two buckets thanks to the overflow metadata. 
%}

%\textcolor{BrickRed}{
%PFHT~\cite{PFHT}, path hashing~\cite{PathHashing} and level hashing~\cite{LevelHashing} are all static hashing schemes. 
%Expanding or shrinking the hash table requires a full-table rehashing (or 1/3 table in level hashing), which is extremely expensive on PM. 
\textbf{Dynamic Hashing on PM.}
CCEH~\cite{CCEH} is a crash-consistent extendible hashing scheme which avoids full-table rehashing~\cite{ExtHashing}. 
To improve search efficiency, it bounds its probing length to four cachelines, but this can lead to low load factor and frequent segment splits. 
CCEH's recovery process requires scanning the directory upon restart, thus sacrifies instant recovery. %, which makes its recovery time scales with data size.
%CCEH avoids logging for directory updates.
%While this design improves performance, it requires a full scan of the directory upon recovery, sacrificing instant recovery.
%\dash's bucket load-balancing techniques improve load factor and reduce segment splits, allowing it to use lightweight logging for directory updates and .
%\dash leverages lazy recovery to reduce service downtime and achieve instant recovery. 
%Concurrency was rarely discussed in detail in prior PM hashing proposals.
Prior proposals often use pessimistic locking~\cite{LevelHashing,CCEH} which can easily become a bottleneck due to excessive PM writes when manipulating locks.
The result is even conflict-free search operations cannot scale.
NVC-hashmap~\cite{NVCHashmap} is a lock-free, persistent hash table based on split-ordered lists~\cite{SplitOrderedLists}.  
Although the lock-free design can reduce PM writes, it is hard to implement; the linked list design may also incur many cache misses.
\dash solves these problems with optimistic locking that reduces PM writes and allows near-linear scalability for search operations.%} 
%Moreover, its linked-list structure makes it cache-unfriendly on PM.  

% to achieve persistence. %ease implementwith carefully flushes inserted in important points to ensure persistence.
%Using the idea of split-ordered lists, buckets are pointers to elements in the list instead of containers of elements, allowing the hash table to expand and shrink conveniently.

%\textcolor{BrickRed}{
\textbf{Range Indexes.} 
%Range indexes are also being actively developed for PM. 
Most range indexes for PM are B+-tree or trie variants and aim to reduce PM writes~\cite{Hwang2018,FPTree,BzTree,Chen2015,NV-Tree,DBPCM,PMwCAS,Lee2017,HiKV}. 
An effective technique is unsorted leaf nodes~\cite{Chen2015,DBPCM,NV-Tree,FPTree} at the cost of linear scans, while hash indexes mainly reduce PM writes by avoiding consecutive displacements.
FP-tree\cite{FPTree} proposes fingerprints in leaf nodes to reduce PM accesses; \dash adopted it to reduce unnecessary bucket probing and efficiently support variable-length keys.
Some work~\cite{FPTree,HiKV} places part of the index in DRAM (e.g., inner nodes) to improve performance. % and the rest (essential) in PM.
This trades off instant recovery as the DRAM part has to be rebuilt upon restart~\cite{PiBench}. %, so recovery time will scale with data size~\cite{PiBench}.
%The hybrid design is suitable for tree-based data structures but hard to be applied to the hash indexes because of the flat structure of the hash table. 
The same tradeoff can be seen in hash tables by placing the directory in DRAM. % to avoid hitting PM on directory entries.
%Although we could put the directory of dynamic hashing (e.g., extendible hashing) in DRAM, it needs to rebuild the directory during recovery, which makes instant recovery impossible. 
%The downside of placing the directory in PM is performance may degrade due to cache misses when accessing directory entries.
%\dash uses large segments to alleviate this problem: the directory is usually small enough to be fit in CPU caches. 
With bucket load balancing techniques, \dash can use larger segments and place the directory in PM, avoiding this tradeoff.%}

%pmdk, allocators (nvm malloc, makalu, pmdk allocator), pm dax file systems (split fs, nova, linux kernel dax)
%\textcolor{BrickRed}{
\textbf{PM Programming.} 
%PM programming requires support at both the OS and userspace levels.
%PM programming framework is needed for the correctness of persistent data structures (e.g., avoiding PM leaks, correct recovery) and also impacts their performance. 
PM data structures rely heavily on userspace libraries and OS support to easily handle such issues as PM allocation and space management. 
PMDK~\cite{PMDK} is so far the most popular and comprehensive library.
%\dash usesbut with several simple changes to allow easier implementation using 8-byte pointers.
An important issue in these libraries is to avoid leaking PM permanently. %PM's byte-addressability and persistence require developers to carefully avoid permanently leaking PM. 
A common solution~\cite{nvmmalloc,Makalu,PMDK,PAllocator} is to use an allocate-activate approach so that the allocated PM is either owned by the application or the allocator upon a crash. % that exhibits an interface similar to \texttt{posix\_memalign}~\cite{POSIX} to guarantee safe transfer of memory ownership.
%Makalu~\cite{Makalu} achieves faster small-block allocations by only enforcing the persistence of critical metadata. %, and rebuilds the rest during recovery. 
%PAllocator~\cite{PAllocator} identifies tackles the challenge of PM fragmentation.
%Although PM allocators are slower, we find the impact on hash tables is not as large as in data structures that use copy-on-write~\cite{PiBench}.
At the OS level, PM file systems provide direct access (DAX) to bypass caches and allow pointer-based accesses~\cite{ext4}. %~\cite{BPFS,PMFS,NOVA,Strata,Aerie,splitfs,ext4}.
%The Linux ext4 file system~\cite{ext4} proposes the DAX mode to enable the direct access to PM from applications. 
Some traditional file systems (e.g., ext4 and XFS) have been adapted to support DAX.
PM-specific file systems are also being proposed to further reduce overhead~\cite{NOVA,BPFS,PMFS,Strata,Aerie,splitfs}. % by redesigning in-kernel file system~\cite{NOVA,BPFS,PMFS} or accessing PM from user-space~\cite{Strata,Aerie,splitfs}. 
We find support for PM programming is still in its early stage and evolving quickly with possible bugs and inefficiencies as Section~\ref{subsec:pm-infra} shows.
This requires careful integration and testing when designing future PM data structures.

\section{Conclusion}
\label{sec:conclusion}
Persistent memory brings new challenges to persistent hash tables in both performance (scalability) and functionality.
We identify that the key is to reduce both unnecessary PM reads and writes, whereas prior work solely focused on reducing PM writes and ignored many practical issues such as PM management and concurrency control, and traded off instant recovery capability.
Our solution is \dash, a holistic approach to scalable PM hashing.
\dash combines both new and existing techniques, including (1) fingerprinting to reduce PM accesses, (2) optimistic locking, and (3) a novel bucket load balancing technique. 
Using \dash, we adapted extendible hashing and linear hashing to work on PM.
On real Intel Optane DCPMM, \dash scales with up to $\sim$3.9$\times$ better performance than prior state-of-the-art, while maintaining desirable properties, including high load factor and sub-second level instant recovery.

%\section*{\bf\small Acknowledgements}
\noindent\textbf{\\Acknowledgements.} We thank Lucas Lersch (TU Dresden \& SAP) and Eric Munson (University of Toronto) for helping us isolate bugs in the
Linux kernel.
We also thank the anonymous reviewers and shepherd for their constructive feedback, and the PC chair's coordination in the shepherding process.
This work is partially supported by an NSERC Discovery Grant, Hong Kong General Research Fund (14200817, 15200715, 15204116), Hong Kong AoE/P-404/18, Innovation and Technology Fund ITS/310/18.

\newpage
\balance
\bibliographystyle{abbrv}
\bibliography{ref}

\end{document}